\documentclass[twocolumn,aps,prx,groupedaddress,longbibliography]{revtex4-1}
\usepackage{graphicx}
\usepackage{color}
\usepackage{amsmath}
\usepackage{marvosym}
\usepackage{bm}

\usepackage{amsthm}

\newcommand{\be}{\begin{equation}}
\newcommand{\ee}{\end{equation}}
\newcommand{\bea}{\begin{eqnarray}}
\newcommand{\eea}{\end{eqnarray}}

\newcommand{\la}{\langle}
\newcommand{\ra}{\rangle}

\newcommand{\bk}{{\bf k}}
\newcommand{\Tr}{{\rm \, Tr\,}}

\newcommand{\ket}[1]{| #1\ra}
\newcommand{\Bra}[1]{\la #1|}
\newcommand{\Ket}[1]{| #1\ra}
\newcommand{\Avg}[1]{\langle #1 \rangle}
\newcommand{\MatEl}[3]{\langle #1|#2|#3\rangle}
\newcommand{\Amp}[2]{\langle #1|#2\rangle}

\renewcommand{\vec}[1]{{\bf #1}}
\renewcommand{\epsilon}{\varepsilon}

\def \bA {{\mathbf{A}}}
\def \bE {{\mathbf{E}}}
\def \bJ {{\mathbf{J}}}

\makeatletter
\def\l@subsubsection#1#2{}
\makeatother

\begin{document}

\title{The Floquet Engineer's Handbook} 
\author{Mark S. Rudner$^1$ and Netanel H. Lindner$^2$}
\affiliation{$^{1}$Niels Bohr International Academy and the Center for Quantum Devices, Niels Bohr
Institute, University of Copenhagen, 2100 Copenhagen, Denmark} \affiliation{$^{2}$Physics Department, Technion, 320003, Haifa, Israel}

\begin{abstract}
We provide a pedagogical technical guide to many of the key theoretical tools and ideas that underlie work in the field of Floquet engineering.
We hope that this document will serve as a useful resource for new researchers aiming to enter the field, as well as experienced researchers who wish to gain new insight into familiar or possibly unfamiliar methods.
This guide was developed out of supplementary material as a companion to our recent review, ``Band structure engineering and non-equilibrium dynamics in Floquet topological insulators,'' Nature Reviews Physics {\bf 2}, 229 (2020). 
The primary focus is on analytical techniques relevant for Floquet-Bloch band engineering and related many-body dynamics.
We will continue to update this document over time to include additional content, and welcome suggestions for further topics to consider.

\end{abstract}

\maketitle

\tableofcontents

\section{Introduction}

Time-periodic driving is a powerful tool for controlling quantum few- and many-body systems, with the potential to enable ``on-demand'' dynamical control of material properties~\cite{Yao2007, Oka2009, Kitagawa2010, Lindner2011, Basov2017}.
Periodic driving has been explored, for example, as a means to open and modify band gaps in graphene and topological insulator surfaces~\cite{Oka2009, Gu2011, Kitagawa2011, Delplace2013, Usaj2014, FoaTorresMultiTerminal, Dehghani2014, Dehghani2015, Sentef2015, Wang2013, Jotzu2014, Aidelsburger2015, Flaschner2016, Tarnowski2019, Asteria2019, McIver2018}, to create topologically non-trivial bands in topologically trivial systems~\cite{Jaksch2003, Mueller2004, Sorensen2005, Oka2009, Kitagawa2010, Lindner2011, Jiang2011, Lindner2013, GomezLeon2013, Wang2014, Chan2016a, Chan2016b, Huebener2017, Thakurathi2017, Kennes2018b, Aidelsburger2015, Lubatsch2019a}, and to access entirely new types of non-equilibrium phases of matter with novel properties that may not exist in equilibrium~\cite{Kitagawa2011, Jiang2011, Rudner2013, AFAI, Benito2014, Reichl2014, Khemani2016, Po2016, vonKeyserlingk2016, Else2016, Potter2016, Harper2017, Else2016b, Hu2015, Quelle2017, Liu2019, Mukherjee2017, Maczewsky2017, Cheng2019, Wintersperger2020}.
These applications and various aspects of the field have been summarized in a number of reviews~\cite{Cayssol2013, BukovReview, Moessner2017, Eckardt2017, Cooper2019, Oka2019, Harper2019, RudnerLindnerReview}.

The aim of this ``Floquet engineer's handbook'' is to provide a pedagogical technical guide to many of the key theoretical tools and ideas that underlie work in this field.
We hope that it will serve as a useful resource for new researchers aiming to enter the field, as well as experienced researchers who wish to gain new insight into familiar or possibly unfamiliar methods.
As indicated above, the field is very broad and encompasses a wide variety of systems and phenomena.
As such, a wide range of techniques are commonly employed.
We primarily focus on analytical techniques relevant for Floquet-Bloch band engineering and related many-body dynamics. 

The topics covered below are organized as follows.
In Sec.~\ref{sec:FloquetBasics} we review the basics of Floquet theory, applied to the time-evolution of periodically-driven quantum systems.
We discuss the transformation to Fourier space, where the problem of time-evolution with a time-dependent Hamiltonian is recast as a problem of matrix diagonalization in an enlarged space.
We further discuss how this approach can be helpful for numerically-obtaining approximate Floquet states, and as a basis for analytical perturbative treatments.
Then in Sec.~\ref{sec:GapOpening} we present a simple explicit calculation that demonstrates how a circularly polarized driving field (analogous to circularly polarized light applied at normal incidence) induces a gap opening in a two-dimensional Dirac system such as graphene. 
In Sec.~\ref{sec:WindingVanish} we briefly recount an important argument about a constraint on the winding numbers of Floquet-Bloch bands. 
In Sec.~\ref{sec:TimeAvgSpec} we introduce and discuss an important object in Floquet analysis: the time-averaged spectral function.
The time-averaged spectral function provides a useful way of ``unfolding'' the single-particle Floquet spectrum.
This quantity allows a meaningful connection  to be drawn between the quasienergies of Floquet states in a driven system and the energies of states in the system's (non-driven) surroundings.
Finally, in Sec.~\ref{sec:FloquetKubo} we discuss the Floquet-Kubo formalism for describing the linear response properties of periodically-driven systems, and the conditions under which a many-body system with Floquet-engineered bands may be expected to exhibit response characteristics similar to those exhibited by a non-driven system with an analogous band structure. 

\section{Floquet theory basics and formalism}\label{sec:FloquetBasics}
Floquet's theorem provides a powerful framework for analyzing periodically-driven quantum systems.
For a system with evolution governed by a time-periodic Hamiltonian $H(t)$ with driving period $T$,
\be
\label{eq:TDSE}  i\hbar \frac{d}{dt}\Ket{\psi(t)} = H(t) \Ket{\psi(t)}, \quad H(t + T) = H(t),
\ee
Floquet states are stationary states of the stroboscopic (``Floquet'') evolution operator
\be
\label{eq:U(T)} U(T) = \mathcal{T}e^{-(i/\hbar)\int_0^T dt'\, H(t')}.
\ee
Here $\mathcal{T}$ denotes time ordering.
The operator $U(T)$ propagates the system forward in time through one complete period of the drive.
Once found, the Floquet states and their associated quasienergies, defined and discussed in detail below,  may provide a useful basis for describing the dynamics of the system.
In this section we discuss the ``extended space'' formalism, which both provides an efficient scheme for numerically obtaining Floquet states and a basis for further theoretical analysis including the application of perturbative techniques.


\subsection{Frequency-space formulation}
A conceptually natural approach to obtaining Floquet states is to obtain (and then diagonalize) the stroboscopic Floquet evolution operator $U(T)$ in Eq.~(\ref{eq:U(T)}) through ``brute force'' direct integration of the time-dependent Schr\"{o}dinger equation (\ref{eq:TDSE}). 
Alternatively, as we now discuss, we may use the structure of the Floquet states, as described by Floquet's theorem, to convert the problem of time-integration to a problem of diagonalization of a single associated Hermitian operator defined on an enlarged space (compared to the Hilbert space of the original problem).

In the context of time evolution of quantum systems with time-periodic Hamiltonians, Floquet's theorem~\cite{Floquet1883} can be stated as follows:\\

\noindent {\bf Floquet's Theorem} -- Consider a quantum system with dynamics governed by a time-periodic Hamiltonian $H(t + T) = H(t)$, with driving period $T$.
The system's evolution can be expanded in a complete basis of orthonormal quasi-stationary ``Floquet states'' $\{\Ket{\psi_n(t)}\}$, which under the time-evolution generated by $H(t)$ satisfy $\Ket{\psi_n(t + T)} = e^{-i\epsilon_n T/\hbar}\Ket{\psi_n(t)}$.
The parameter $\epsilon_n$ is the ``quasienergy'' of Floquet state $\Ket{\psi_n(t)}$.\\

Bloch's theorem, which is widely familiar in solid state physics, is another special case of Floquet's theorem, applied to the eigenstates of an electron in a 
crystal with a potential that has a periodic structure in space.
Analogous to how a Bloch state can be decomposed into a product of a plane wave and a periodic function (with the same periodicity as the underlying lattice), each Floquet state can be decomposed as:
\begin{equation}
\label{eq:FloquetDecomp} \Ket{\psi_n(t)} = e^{-i \epsilon_n t/\hbar}\Ket{\Phi_n(t)}, \quad \Ket{\Phi_n(t + T)} = \Ket{\Phi_n(t)}.
\end{equation}
Importantly, due to the fact that $\Ket{\Phi_n(t)}$ exhibits the same time-periodicity as the drive, it can be represented in terms of a discrete Fourier series in terms of harmonics of the drive frequency, $\omega = 2\pi/T$:
\begin{equation}
\label{eq:FloquetExpansion} \Ket{\Phi_n(t)} = \sum_m e^{-i m \omega t} \Ket{\phi_n^{(m)}},
\end{equation}
where $\Ket{\phi_n^{(m)}}$ is the $m$-th Fourier coefficient of $\Ket{\Phi_n(t)}$.
Note that {i) the Fourier coefficients $\Ket{\phi_n^{(m)}}$ are not normalized, i.e., generically $\Amp{\phi_n^{(m)}}{\phi_n^{(m)}} < 1$, and ii) there is no simple orthogonality relation between different Fourier coefficients $\Ket{\phi_n^{(m)}}$ and $\Ket{\phi_n^{(m')}}$,} apart from the (nontrivial) fact that they must add up to produce orthogonal Floquet states $\Ket{\psi_n(t)}$ at {\it every} time, $t$.

Our goal is now to find an {\it algebraic} equation that yields the Fourier coefficients $\{\Ket{\phi_n^{(m)}}\}$ for the Floquet state solutions to the Schr\"{o}dinger equation (\ref{eq:TDSE}). 
Substituting Eq.~(\ref{eq:FloquetDecomp}) on both sides of Eq.~(\ref{eq:TDSE}) and canceling factors of $e^{-i \epsilon_n t/\hbar}$ yields
\begin{equation}
 \label{eq:EZDeriv1} \left[\epsilon_n + i \hbar \frac{d}{dt}\right]\Ket{\Phi_n(t)} = H(t) \Ket{\Phi_n(t)}.
\end{equation}
Using the Fourier decomposition of $\Ket{\Phi_n(t)}$ in Eq.~(\ref{eq:FloquetExpansion}) and $H(t) = \sum_m e^{-i m \omega t} H^{(m)}$, which follows from the periodicity of $H(t)$, Eq.~(\ref{eq:EZDeriv1}) yields:
\be
 \label{eq:EZDeriv2} \left(\epsilon_n + m\hbar\omega\right)\Ket{\phi^{(m)}_n} = \sum_{m'}H^{(m - m')} \Ket{\phi^{(m')}_n}.
\ee

As a final step, we express Eq.~(\ref{eq:EZDeriv2}) as an eigenvalue equation in Fourier harmonic space.
Specifically, we create a vector $\bm{\varphi}_n$ by ``stacking up'' the Fourier coefficients $\{\Ket{\phi^{(m)}_n}\}$, and rearrange the coefficients in Eq.~(\ref{eq:EZDeriv2}) into a matrix $\mathcal{H}$ acting on vectors in this space: 
\begin{widetext}
\bea
\label{eq:EZFloquet}\mathcal{H} \bm{\varphi}_n = \epsilon_n \bm{\varphi}_n, \quad
\mathcal{H} =
\left(\begin{array}{ccccc}
\ddots & H^{(-1)} & H^{(-2)} &  \\
H^{(1)} & H_0 - m \hbar \omega & H^{(-1)} &  H^{(-2)} \\
H^{(2)} & H^{(1)} & H_0 - (m+1) \hbar \omega & H^{(-1)} \\
&  H^{(2)} & H^{(1)} &   \ddots
\end{array}\right), \quad
\bm{\varphi}_n =  \left(\begin{array}{c}
\vdots  \\ \Ket{\phi^{(m)}_n} \\ \Ket{\phi^{(m+1)}_n} \\ \vdots
\end{array}\right).
\eea
\end{widetext}
Here, $H_0 \equiv H^{(0)} = \frac{1}{T} \int_0^T dt\, H(t)$ is the dc (time-averaged) part of the Hamiltonian.
Note that the matrix $\mathcal{H}$ in Eq.~(\ref{eq:EZFloquet}) has a block structure:
each block is of size $d \times d$, where $d$ is the dimension of the Hilbert space of the system [i.e., the dimension of $H(t)$].
The number of blocks, which are labeled by Fourier harmonic indices, is formally infinite, since the sum over $m$ in Eq.~(\ref{eq:FloquetDecomp}) ranges over all integers.

The time-periodic part of the Floquet state wave function, $\Ket{\Phi_n(t)}$, is obtained from the vector of Fourier coefficients $\bm{\varphi}_n$ through multiplication with a rectangular matrix of oscillatory phase factors, $\mathcal{P}(\omega t)~=~( \cdots\, e^{-i m \omega t}\bm{1}_{d\times d} \ e^{-i(m+1)\omega t}\bm{1}_{d\times d} \, \cdots )$.
Here $\bm{1}_{d\times d}$ stands for the identity operator in the physical Hilbert space, taken to have dimension $d$ (see above).
{Explicitly, this gives}
\bea
\nonumber \Ket{\Phi_n(t)} &=& \mathcal{P}(\omega t)\, \bm{\varphi}_n\\ 
\label{eq:EZ2Phys} &=& \sum_m e^{-i m \omega t}\Ket{\varphi^{(m)}_n}.
\eea
The projector $\mathcal{P}(\omega t)$ thus plays a central role in the mapping between extended space and the physical Hilbert space.\\ 

\noindent{\it Overcompleteness of solutions to Eq.~(\ref{eq:EZFloquet})}.--- As mentioned above, the dimension of the matrix $\mathcal{H}$ in Eq.~(\ref{eq:EZFloquet}) is larger than that of the Hilbert space of the original problem in Eq.~(\ref{eq:TDSE}).
Thus the solutions to Eq.~(\ref{eq:EZFloquet}) must correspond to linearly-dependent quantum states.
In fact, the Fourier space formulation of the Floquet problem encodes a redundancy, such that {\it the same} physical solutions of the form in Eq.~(\ref{eq:FloquetDecomp}) are encoded infinitely-many times in the eigenvectors of $\mathcal{H}$.

\begin{figure}[t]
\includegraphics[width=\columnwidth]{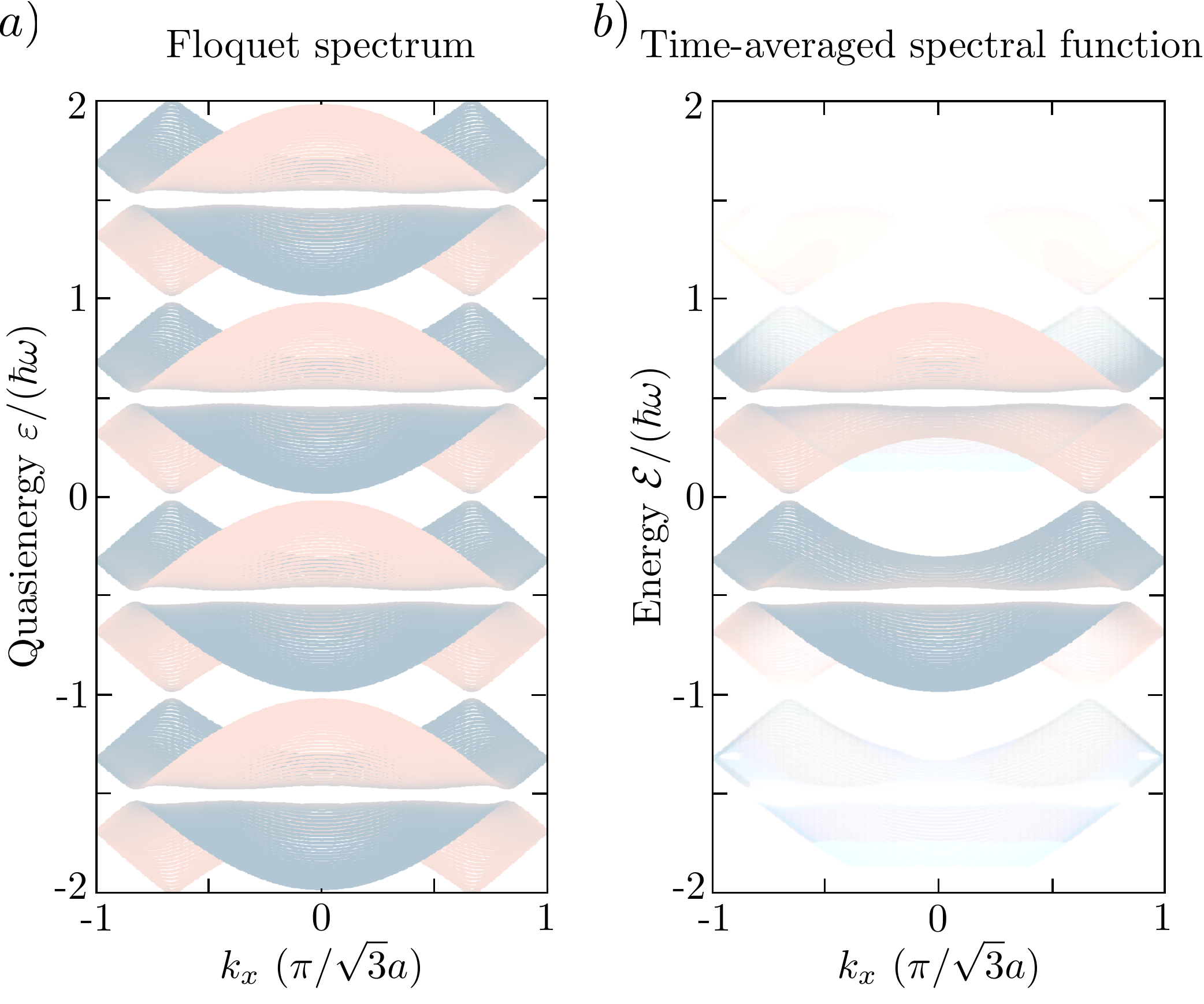}
\caption{Floquet spectrum and time-averaged spectral function.
a) The Floquet (quasienergy) spectrum, obtained by diagonalizing the ``extended-zone'' Floquet Hamiltonian $\mathcal{H}$ in Eq.~(\ref{eq:EZFloquet}), is periodic with a ``Floquet Brillouin zone'' of width equal to the driving field photon energy, $\hbar\omega$.
For illustration, here we show the spectrum of graphene subjected to circularly polarized light.
Colors indicate overlaps of the Floquet states with corresponding valence (blue) and conduction (pink) band states of the system in the absence of the drive.
b) The time-averaged spectral function provides means to ``unfold'' the quasienergy spectrum, and to visualize how the Floquet states of the driven system may couple to external (non-driven) degrees of freedom at specific {\it energies}.
As shown here, for the same system as displayed in panel a), the spectral weights of the sidebands (arising from the original conduction and valence bands being shifted up and down by an integer multiple of the driving field photon energy $\hbar\omega$) decay rapidly with the number of photons absorbed/emitted (indicated by the intensities of the points).
The rates of photon-assisted tunneling between an external lead and the Floquet system, and for example, of photon-assisted electron-phonon ``Floquet-Umklapp'' scattering, are controlled by these sideband intensities.
}
\label{fig:FloquetSpectrum}
\vspace{-0.2 in}
\end{figure}

To see where the redundancy of solutions to Eq.~(\ref{eq:EZFloquet}) comes from, we 
write a general Floquet state in the form $\Ket{\psi_n(t)} = e^{-i\epsilon_n t/\hbar}\sum_m e^{-im\omega t}\Ket{\phi^{(m)}_n}$ [see  Eqs.~(\ref{eq:FloquetDecomp}) and (\ref{eq:FloquetExpansion})].
Note that, {\it without changing} the state $\Ket{\psi_n(t)}$, its quasienergy can be shifted by any integer multiple of $\hbar \omega$ by adding and subtracting any integer multiple of $i \omega t$ in the pair of exponentials in this expression:
\bea
\nonumber  \Ket{\psi_n(t)} &=& e^{-i(\epsilon_n + m'\hbar \omega)t/\hbar}\sum_m e^{-i(m-m')\omega t}\Ket{\phi^{(m)}_n}\\
\label{eq:FloquetRedundancy}&\equiv& e^{-i\tilde{\epsilon}_n t/\hbar} \sum_m e^{-i m \omega t}\Ket{\tilde{\phi}^{(m)}_n}, 
\eea
where $\tilde{\epsilon}_n = \epsilon_n + m' \hbar\omega$ and $\Ket{\tilde{\phi}^{(m)}_n} = \Ket{\phi^{(m + m')}_n}$.
Thus {\it all distinct Floquet state solutions} to the Schr\"{o}dinger equation can be indexed with quasienergies that fall within a single ``Floquet-Brillouin zone'' of width $\hbar\omega$, {$\epsilon_{\rm min} \le \epsilon < \epsilon_{\rm min} + \hbar\omega$}.
Any solution to Eq.~(\ref{eq:EZFloquet}) with quasienergy outside of the Floquet-Brillouin zone encodes [via Eq.~(\ref{eq:FloquetRedundancy})] an identical physical state to one corresponding to a solution of Eq.~(\ref{eq:EZFloquet}) with quasienergy within the Floquet-Brillouin zone.
Consequently, as illustrated in Fig.~\ref{fig:FloquetSpectrum}a, the spectrum of $\mathcal{H}$ in Eq.~(\ref{eq:EZFloquet}) consists of an infinite number of copies of the system's Floquet spectrum, rigidly shifted up and down by integer multiples of $\hbar\omega$.

At a fundamental level, the redundancy of the Floquet spectrum arises due to the discrete time-translation invariance of the system.
The Floquet states are eigenstates of the stroboscopic evolution (discrete time-translation) operator, $U(T)$, in Eq.~(\ref{eq:U(T)}).
Due to the unitarity of $U(T)$, its eigenvalues are all unit-modulus complex numbers: $U(T)\Ket{\psi_n(t)} = e^{-i\theta_n}\Ket{\psi_n(t)}$, where $\theta_n$ is real.
By parametrizing the eigenvalues of $U(T)$ in terms of quasienergies, $\theta_n = \epsilon_n T / \hbar$, we introduce an artificial multi-valuedness: quasienergy values separated by an integer multiple of $\hbar\omega$ correspond to the {\it same} eigenvalue $e^{-i\epsilon_n T / \hbar}$ of $U(T)$.
In physical terms, the periodicity of quasienergy expresses the fact that, in the presence of a periodic drive, the system's energy is conserved up to the absorption or emission of integer multiples of the driving field photon energy, $\hbar\omega$.\\


\subsection{Truncation of frequency space}\label{sec:Trunc}
In the previous subsection, we described how the problem of finding Floquet state solutions of the time-dependent Schr\"{o}dinger equation for a periodically-driven system, Eq.~(\ref{eq:TDSE}), can be recast from one of computing a time-ordered exponential [Eq.~(\ref{eq:U(T)})], to one of finding eigenvalues of an enlarged ``extended-zone'' or ``extended-space'' Floquet Hamiltonian, $\mathcal{H}$ [Eq.~(\ref{eq:EZFloquet})].
Given that $\mathcal{H}$ is an infinite-dimensional matrix, even for a system with a finite-dimensional Hilbert space, it may not be clear what we have gained through this transformation.
Fortunately, as we now discuss, $\mathcal{H}$ has a particular structure that can be exploited to enable approximate solutions to Eq.~(\ref{eq:EZFloquet}) to be found efficiently.
This structure moreover has a clear physical interpretation, which may help provide further insight into the dynamics of the system.

To most clearly exhibit the key features of the extended-zone Floquet Hamiltonian, we will focus for now on the case of a single harmonic drive:
\be
\label{eq:HarmonicDrive} H(t) = H_0 + V(t),\quad V(t) = Ve^{i\omega t} + V^\dagger e^{-i\omega t}.
\ee
The corresponding extended-zone Floquet Hamiltonian $\mathcal{H}$ then takes the form in Eq.~(\ref{eq:EZFloquet}) with $H^{(-1)} = V$, $H^{(1)} = V^\dagger$, and $H^{(\Delta m)} = 0$ for $|\Delta m| > 1$.
In this case, $\mathcal{H}$ has a block-tridiagonal form, with the non-driven Hamiltonian $H_0$ (shifted by multiples of $\hbar\omega$) copied over and over in the blocks along the diagonal, and the drive coupling $V$ ($V^\dagger$) placed in the blocks above (below) the diagonal:
\be
\label{eq:H0V}\mathcal{H} =
\left(\begin{array}{cccc}
\ddots &V & 0 \\ 
V^\dagger & H_0 - m \hbar \omega & V \\ 
0 & V^\dagger & \ddots 
\end{array}\right).
\ee

\begin{figure}[t]
\includegraphics[width=\columnwidth]{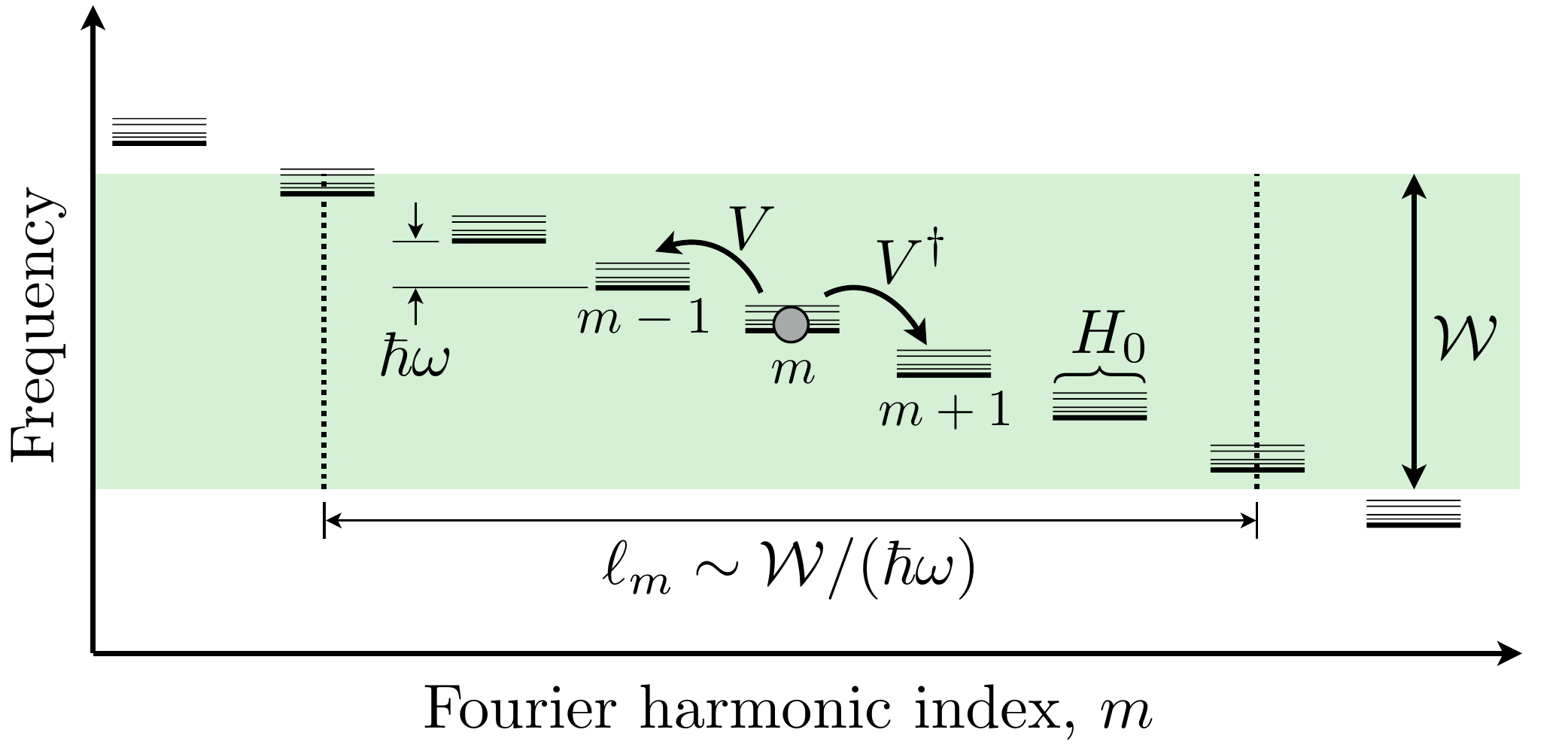}
\caption{The extended-zone Floquet Hamiltonian, Eq.~(\ref{eq:EZFloquet}), can be interpreted as describing the one-dimensional tight-binding dynamics of an effective particle hopping in Fourier harmonic space.
The site energies and couplings within each unit cell, $m$, are described by the time-averaged Hamiltonian, $H_0$.
For a drive with a single harmonic, as described by Eqs.~(\ref{eq:HarmonicDrive}) and (\ref{eq:H0V}), the drive $V$ induces hopping between neighboring unit cells, in which the Fourier harmonic shifts by one.
The on-site energies vary linearly with $m$ due to the shifts $-m\hbar\omega$ appearing in the diagonal blocks of Eqs.~(\ref{eq:EZFloquet}) and (\ref{eq:H0V}).
This linear potential acts as a uniform electric field on the effective particle.
In the absence of the linear potential, the eigenstates would be Bloch waves in $m$-space, with corresponding total bandwidth $\mathcal{W} \sim {\rm max}\{||V||, ||H_0||\}$.
From a semiclassical point of view, in the presence of the linear potential the effective particle may explore a range of harmonics of width $\ell_m \sim \mathcal{W}/(\hbar\omega)$, where its kinetic and potential energies may be balanced; the Floquet state wave functions are therefore {\it localized} in harmonic space and the infinite-dimensional matrices in Eqs.~(\ref{eq:EZFloquet}) and (\ref{eq:H0V}) can be safely truncated. 
}
\label{fig:effective_hopping}
\vspace{-0.2 in}
\end{figure}

Importantly, the (block) tridiagonal structure of $\mathcal{H}$ is closely analogous to that of a tight-binding lattice with nearest-neighbor hopping and a linear potential (i.e., to the problem of an electron in a one-dimensional lattice with a uniform electric field), see Fig.~\ref{fig:effective_hopping}.
By making analogies to Bloch oscillations~\cite{Bloch1929, Landau1932, Zener1934} and to the ``Wannier-Stark ladder''~\cite{Wannier1960}, we may gain important insight into the nature of the Floquet states encoded in the solutions to Eq.~(\ref{eq:EZFloquet}).
These insights will furthermore shed light on the utility of this approach as a numerical method for obtaining Floquet states.

Consider the situation where the system described by $H_0$ has a finite dimensional Hilbert space (e.g., for a two-band Floquet-Bloch system, $H_0$ could be the $2 \times 2$ Bloch Hamiltonian describing the dynamics for a given crystal momentum, $\vec{k}$).
Neglecting the ``linear potential'' $-m\hbar\omega$ in Eq.~(\ref{eq:H0V}), we interpret the matrix $\mathcal{H}$ as the Hamiltonian of an effective particle hopping between sites/unit cells of a one-dimensional lattice in Fourier harmonic space.
The on-site energies and intra-unit-cell hoppings are described by $H_0$, while the inter-unit-cell hopping is described by $V$, see Fig.~\ref{fig:effective_hopping}.
In particular, for a system where $H_0$ has dimension one, the kinetic energy of the effective particle has a bandwidth $\mathcal{W} = 2|V|$.
Crucially, even when the dimension of $H_0$ is larger than one (but finite), the effective particle system (in the absence of the $m\hbar\omega$ terms) still has a finite bandwidth, $\mathcal{W}$, approximately given by the maximum between the norms $||V||$ and $||H_0||$.

In the absence of the linear potential $-m\hbar\omega$, the eigenstates of the effective particle are extended Bloch waves in the Fourier harmonic space.
These eigenstates become {\it localized} in Fourier harmonic space when the $-m\hbar\omega$ ``potential energy'' terms are introduced.
From a semiclassical point of view, this localization arises from the fact that, due to the finite bandwidth of the effective particle's kinetic energy, the total energy given by the sum of potential and kinetic energies can only take a constant value over a finite range of $m$-values, $\ell_m$, see Fig.~\ref{fig:effective_hopping}.
The range $\ell_m$ of Fourier coefficients accessible at a given quasienergy $\epsilon$ can be estimated as $\ell_m \sim \mathcal{W}/(\hbar\omega)$, with the Fourier-space wave function falling off (faster than) exponentially in $m$ outside of this ``classically-allowed'' region~\cite{Lindner2017}.

Importantly, due to the fact that the each eigenvector of the infinite-dimensional matrix $\mathcal{H}$ in Eq.~(\ref{eq:H0V}) has support only over a limited range of Fourier harmonics, $m$, a complete set of Floquet states with quasienergies in one Floquet-Brillouin zone can be found to arbitrary accuracy by using a {\it truncated}, finite-dimensional sub-matrix of $\mathcal{H}$, which we call $\mathcal{H}_{\rm trunc}$.
In particular, by restricting the set of Fourier coefficients to a range of $(2M + 1)$ values of width much larger than the localization radius of the Fourier-space wave functions, $(2M + 1) \gg \ell_m$, the difference between the true Floquet state wave functions and the approximate wave functions obtained as eigenvectors of the truncated extended-zone Floquet Hamiltonian can be made exponentially small.
Importantly, the truncation of $\mathcal{H}$ breaks the perfect repetition/redundancy of the Floquet spectrum between different Floquet-Brillouin zones.
For Floquet-zones centered at $m$-values near the truncation boundaries, the spectrum and eigenstates may be severely distorted.
However, as long as the truncation range is taken to be large enough, the eigenvectors and eigenvalues near the middle of the spectrum of $\mathcal{H}_{\rm trunc}$ can be used to numerically obtain approximate Floquet states with an error that decays exponentially with $M$.\\


\noindent {\it Relation to system coupled to quantized harmonic mode}.---
Despite the fact that Eqs.~(\ref{eq:TDSE}), (\ref{eq:EZFloquet})  and (\ref{eq:H0V}) 
describe a system subjected to a classical time-dependent drive, the structure of the extended-zone Floquet Hamiltonian $\mathcal{H}$ is similar to that of the same system with the drive replaced by a quantized harmonic oscillator mode.
In this analogy, the Fourier harmonic indices label Fock states of the oscillator mode. 
The diagonal blocks, which take the form $H_0 - m\hbar\omega$, describe the system together with the oscillator, where $m$ photons have been removed from the oscillator state relative to an arbitrary reference Fock state (which must be assumed to contain a large number of photons).
The off-diagonal blocks $H^{(\Delta m)}$ describe couplings in which $\Delta m$ photons are absorbed ($\Delta m > 0$) or emitted ($\Delta m < 0$) by the system.

While the similarities mentioned above may be useful for gaining conceptual insight into the Floquet problem, there are several key differences between $\mathcal{H}$ and a true system-oscillator coupling Hamiltonian.
First, recall that, when applied to a Fock state, the creation and annihilation operators $a^\dagger$ and $a$ of a harmonic mode generate factors proportional to the square root of the number of photons present: $a\Ket{n}_\omega = \sqrt{n}\Ket{n-1}_\omega$, where $\Ket{n}_\omega$ labels a Fock state with $n$ photons in the mode with frequency $\omega$.
Therefore, in a true system-oscillator Hamiltonian, the off-diagonal blocks $H^{(\Delta m)}$ would contain an explicit dependence on the photon number index; in contrast, the off-diagonal blocks in $\mathcal{H}$ are translation-invariant, depending only on the net number of photons absorbed, $\Delta m$.
Second, while a true oscillator mode has a vacuum state $\Ket{n = 0}_\omega$, such that the photon-number space is semi-infinite (with the photon number index ranging only over non-negative integers), the Fourier harmonic index labeling the blocks of $\mathcal{H}$ runs over all integers.
The system dynamics induced by the classical drive, as described in Eq.~(\ref{eq:TDSE}), can be effectively recovered by placing the oscillator in a large-amplitude coherent state, where the square-root dependencies of the matrix elements may be ignored and where the support of the oscillator wave function is localized far from the vacuum state in photon-number space.


\subsection{Floquet Fermi's golden rule}

In this subsection we derive a Floquet-version of Fermi's golden rule, and show how it can be
evaluated within the extended space formalism. Our aim is to calculate the rate of transitions between (unperturbed) Floquet states, induced by a weak perturbation. Of
particular relevance for FTIs, as an application we will use Floquet Fermi's golden rule in the
extended space to gain insight into the nature of Floquet-Umklapp  scattering processes (as discussed in Sec.~V of the Ref.~\cite{RudnerLindnerReview} and, e.g.,  Refs.~\cite{Dehghani2014, Seetharam2015, Iadecola2015, Bilitewski2015a, Genske2015, Esin2018, Seetharam2019}).


Consider a periodically-driven system, with Hamiltonian $H(t) = H(t + T)$, subjected to a weak
perturbation $H_{\rm int}$. 
{For simplicity, here} we consider the case where $H_{\rm int}$ is a static (time-independent)
perturbation.  
The time-evolution of the system is governed
by $i\hbar\frac{d}{dt}\Ket{\psi(t)} = [H(t) + H_{\rm int}]\Ket{\psi(t)}$. We emphasize that the
drive itself may be strong: only the additional perturbation $H_{\rm int}$ is taken to be weak.

As a starting point, we assume that a complete set of Floquet states of the unperturbed problem (with $H_{\rm int} = 0$) are known.
We denote these unperturbed Floquet states as $\{\Ket{\psi_n(t)}\}$, with corresponding quasienergies $\{\epsilon_n\}$.
As discussed above, a complete set of Floquet states can be indexed within a single Floquet-Brillouin zone.
Therefore, we take all $\{\epsilon_n\}$ to satisfy $\epsilon_{\rm min} \le \epsilon_n < \epsilon_{\rm min} + \hbar\omega$.

Suppose that the system is in the unperturbed Floquet state $\Ket{\psi_{i}(0)}$ at time $t = 0$.
We calculate the probability that, due to the perturbation $H_{\rm int}$, the system has made a transition to unperturbed Floquet state $\Ket{\psi_f(NT)}$ after $N$ periods of the drive, where $N$ is an integer.
After time $NT$, the transition probability $P_{fi}(NT)$ from $\Ket{\psi_i(0)}$ to (unperturbed) 
Floquet state $\Ket{\psi_f(NT)}$ is given by
 \be
 P_{fi}(NT) = |\Amp{\psi_f(NT)}{\psi(NT)}|^2,
 \label{eq: Pfi def}
 \ee
where $\Ket{\psi(NT)}$ is the state of the system evolved in the presence of the perturbation. 
Below we will see that, by focusing on these stroboscopic probabilities, we will obtain a simple expression for the transition rate in terms of the Fourier component vectors $\bm{\varphi}_i$ and $\bm{\varphi}_f$ in extended space.
We will further show that these results directly extend to times that are not necessarily integer multiples of the driving period. 

To obtain the {\it rate} of transitions, i.e., of linear growth of the $P_{fi}(NT)$ with $NT$, we evaluate the transition probability $P^{(1)}_{fi}(NT)$ using the evolution expanded to first order in $H_{\rm int}$.
As a first step, we move to the interaction picture $\Ket{\psi(t)}_I =
U^\dagger(t, 0)\Ket{\psi(t)}$, where $U(t, 0) = \mathcal{T}e^{-(i/\hbar)\int_0^t dt' H(t')}$ is the
evolution operator of the unperturbed system. In this interaction picture, the Floquet states
$\{\Ket{\psi_n(t)}\}$ are stationary: $\Ket{\psi_n(t)}_I = \Ket{\psi_n(0)}$.

Writing the overlap in Eq.~(\ref{eq: Pfi def}) using the interaction picture states and evolution operator, and expanding to first order in $H_{{\rm int},I}(t) = U^\dagger(t, 0)\, H_{{\rm int}}\, U(t, 0)$, 
we obtain 
\be
 \label{eq:P_def}P^{(1)}_{fi}(NT)= \left|\frac{-i}{\hbar}\int_0^{NT} \!\!\!dt\, \MatEl{\psi_f(0)}{H_{{\rm int},I}(t)}{\psi_i(0)}\right|^2.
\ee
Here 
we used $\Ket{\psi_f(NT)}_I = \Ket{\psi_f(0)}$.
The corresponding transition rate, $W_{fi}$, is then given by
\be
\label{eq:W_def} W_{fi} = \lim_{N\to\infty} \frac{1}{NT} P^{(1)}_{fi}(NT).
\ee
Below we derive an exact expression for $P^{(1)}_{fi}(NT)$, and then obtain $W_{fi}$ by taking the limit in Eq.~(\ref{eq:W_def}).

Returning to Eq.~(\ref{eq:P_def}), we apply the evolution operators $U(t, 0)$ and $U^\dagger(t, 0)$ in $H_{{\rm int},I}(t)$ to the right and left, respectively.
Writing the Floquet states as $\Ket{\psi_{i, f}(t)} = e^{-i \epsilon_{i,f} t/\hbar}\Ket{\Phi_{i,f}(t)}$, with $\Ket{\Phi_{i,f}(t)} = \Ket{\Phi_{i,f}(t+T)}$, gives
\be
\label{eq:Prob_NT} P^{(1)}_{fi}(NT) = \left|\frac{-i}{\hbar}\int_0^{NT}\!\!\! dt\, e^{i\Delta\epsilon t/\hbar} \MatEl{\Phi_f(t)}{H_{\rm int}}{\Phi_i(t)} \right|^2,
\ee
where we have defined $\Delta\epsilon \equiv \epsilon_f - \epsilon_i$ for brevity.

The integral in Eq.~(\ref{eq:Prob_NT}) can be split into a sum of integrals over each complete period, $nT \le t < (n+1)T$, for $n = (0, 1, \ldots, N-1)$.
Due to the periodicity of $\Ket{\Phi_{i}(t)}$ and $\Ket{\Phi_{f}(t)}$, for each $n$ the corresponding integral is given by $e^{i\Delta\epsilon (nT)/\hbar}\int_0^{T} dt\, e^{i\Delta\epsilon t/\hbar} \MatEl{\Phi_f(t)}{H_{\rm int}}{\Phi_i(t)}$.
We thus express $P^{(1)}_{fi}(NT)$ in terms of the time-averaged matrix element $M_{fi} \equiv \frac{1}{T} \int_0^{T} dt\, e^{i\Delta\epsilon t/\hbar} \MatEl{\Phi_f(t)}{H_{\rm int}}{\Phi_i(t)}$ as
\be
\label{eq:P_fi_double_sum}P^{(1)}_{fi}(NT) = \frac{T^2}{\hbar^2}|M_{fi}|^2\sum_{n,n' = 0}^{N-1} e^{i\Delta\epsilon (n-n')T/\hbar}.
\ee
The factor of $T^2$ in Eq.~(\ref{eq:P_fi_double_sum}) was obtained by multiplying and dividing by $T^2$ in order to express $P^{(1)}_{fi}(NT)$ in terms of the time-averaged matrix element $M_{fi}$.

The sums over $n$ and $n'$ in Eq.~(\ref{eq:P_fi_double_sum}) can be evaluated by changing to sum and difference indices $\bar{n} = n + n'$ and $\Delta n = n - n'$.
Heuristically, if the sum over $\Delta n$ is taken from $-\infty$ to $\infty$, we obtain a periodic array of delta functions (a ``Dirac comb''): $\sum_{\Delta n} e^{i(\epsilon_f - \epsilon_i) \Delta n T/\hbar} \sim \frac{2\pi\hbar}{T}\sum_{k = -\infty}^\infty \delta( \epsilon_f - \epsilon_i - k \hbar \omega)$.
The sum over $\bar{n}$ gives a contribution proportional to $N$.
All together we thus obtain a transition probability $P^{(1)}_{fi}(NT)$ that grows linearly with $NT$, which gives a finite rate in Eq.~(\ref{eq:W_def}).

More rigorously, the sums appearing in Eq.~(\ref{eq:P_fi_double_sum}) can be evaluated exactly,
analytically. For each value of $\bar{n}$, the sum over $\Delta n$ can be expressed in the form of
a finite geometric series with limits that depend on $\bar{n}$.
Defining $x = e^{2i\Delta\theta}$, with $\Delta\theta = \Delta \epsilon T / \hbar$, the sum over $\Delta n$ can be performed using 
\be
  \sum_{n=0}^{k} x^n = \frac{1 - x^{k+1}}{1 - x}, 
\ee
where $k$ depends on $\bar{n}$. 
 Once the sum over $\Delta n$ is evaluated as a function of $\bar{n}$, the sum over $\bar{n}$ can be performed piecewise, for i) $0\le \bar{n} \le N-1$ and ii)  $N \le \bar{n} \le 2N-2$, using the summation formula (see, e.g., Wikipedia) 
\be 
\sum_{n = 0}^M \sin (\Delta\theta + n\Delta\theta) = \frac{\sin\frac{(M+1)\Delta\theta}{2} \sin (\Delta\theta + \frac{M\Delta\theta}{2})}{\sin \frac{\Delta\theta}{2}}. 
\ee
Evaluating the two sums using upper limits $M = N -1$ and $M = N - 2$, for cases i) and ii), respectively, we obtain:
\be \label{eq:double_sum_final}\sum_{n,n'=0}^{N-1} e^{i\Delta\theta \Delta n} =
\frac{\sin \frac{N\Delta\theta}{2} \left(\sin \frac{(N-1)\Delta\theta}{2} + \sin
\frac{(N+1)\Delta\theta}{2} \right)}{\sin \Delta \theta \sin \frac{\Delta \theta}{2}}. 
\ee 

As a function of $\Delta \theta$, the expression on the right hand side of Eq.~(\ref{eq:double_sum_final}) consists of a periodic array of sharp peaks, centered where $\Delta \theta$ is equal to an integer multiple of $2\pi$.
The peaks become narrower and narrower as $N$ becomes large.
Importantly, because we parametrize the initial state and the complete set of possible final states by quasienergies $\epsilon_i$ and $\{\epsilon_f\}$ within a single Floquet Brillouin zone, the quasienergy difference $\epsilon_f - \epsilon_i$ must always be smaller than $\hbar \omega$.
Correspondingly, we have that $|\Delta\theta| < 2\pi$.
As a result, only the peak around $\Delta \theta = 0$ is relevant. 

Due to the fact that $\sum_{n,n'} e^{i\Delta\theta (n-n')}$ is only significant for $\Delta\theta \ll 1$, we expand Eq.~(\ref{eq:double_sum_final}) for small $\Delta\theta$.
This gives, for large $N$ and $\Delta\theta \ll 1$,
\be
\lim_{N\to \infty}\frac{1}{N}\sum_{n,n'} e^{i\Delta\theta (n-n')} = \lim_{N\to \infty}\frac{1}{N}\frac{\sin^2 \frac{N\Delta\theta}{2}}{(\Delta\theta/2)^2}. 
\ee
Substituting this result into Eq.~(\ref{eq:P_fi_double_sum}), using Eq.~(\ref{eq:W_def}), and restoring $\Delta\theta = \Delta\epsilon T / \hbar$, gives
\be
W_{fi} =  \lim_{N\to\infty} \frac{1}{NT}  \frac{\sin^2 \frac{\Delta\epsilon NT}{2\hbar}}{\left(\frac{\Delta\epsilon}{2}\right)^2}|M_{fi}|^2.
\ee
Recognizing precisely the same squared sinc function that appears in typical derivations of Fermi's golden rule (see, e.g., Ref.~\cite{Sakurai}), we finally obtain the transition rate 
\be
\label{eq:W_fi_final} W_{fi} = \frac{2\pi}{\hbar}  |M_{fi}|^2 \delta(\epsilon_f - \epsilon_i).
\ee
Note that in obtaining Eq.~(\ref{eq:W_fi_final}) we have assumed that the perturbation $H_{\rm int}$ is sufficiently weak such that $W_{fi} \ll \omega$. 
This gives a sufficient time window for the peaks of the function in Eq.~(\ref{eq:double_sum_final}) to become sharp on the scale of $\omega$ before the transition probability saturates. 

To evaluate the time-averaged matrix element $M_{fi}$ [see definition above Eq.~(\ref{eq:P_fi_double_sum})], we note that the delta function sets $\Delta\epsilon = 0$.
Thus, expanding $\Ket{\Phi_{i,f}(t)}$ in terms of Fourier harmonics $\{\Ket{\phi^{(m)}_{i,f}}\}$, we obtain
\bea
\nonumber M_{fi} &=& \sum_{mm'}\frac{1}{T}\int_0^T dt \, e^{-i(m-m')\omega t}\MatEl{\phi^{(m')}_f}{H_{\rm int}}{\phi^{(m)}_i}\\
\label{eq:M_fi_hilbert}&=& \sum_m \MatEl{\phi^{(m)}_f}{H_{\rm int}}{\phi^{(m)}_i}.
\eea
Remarkably, the sum over Fourier harmonic components in Eq.~(\ref{eq:M_fi_hilbert}) is precisely given by $M_{fi} = \bm{\varphi}^\dagger_f \mathcal{H}_{\rm int} \bm{\varphi}_i$, where $\bm{\varphi}_i$ and $\bm{\varphi}_f$ are the extended space Fourier component vectors corresponding to Floquet states $\Ket{\psi_i(t)}$ and $\Ket{\psi_f(t)}$ with quasienergies taken in the specified zone $\epsilon_{\rm min} \le \epsilon_{i,f} < \epsilon_{\rm min} + \hbar \omega$, and $\mathcal{H}_{\rm int}$ is the perturbation Hamiltonian in extended space.
Importantly, due to the fact that $H_{\rm int}$ is independent of time, $\mathcal{H}_{\rm int}$ is diagonal in its Fourier component indices (i.e., all off-diagonal blocks are zero).
All together, this leads to the final expression for Floquet Fermi's golden rule, in terms of the extended space perturbation Hamiltonian and eigenvectors:
\be
\label{eq:W_fi} W_{fi} = \frac{2\pi}{\hbar}  |\bm{\varphi}^\dagger_f\, \mathcal{H}_{\rm int}\, \bm{\varphi}_i|^2 \delta(\epsilon_f - \epsilon_i).
\ee
With this expression, transition rates between Floquet states can be calculated straightforwardly from their extended space representations.

Interestingly, we note that Eq.~(\ref{eq:W_fi}) is precisely what one would obtain for Fermi's golden rule based on naively evolving in time in extended space.
Specifically, consider evolution in extended space with respect to an auxiliary time variable, $\tau$:
\begin{equation}
\label{eq:EZEvolution}i\hbar\frac{d}{d\tau} \boldsymbol{\Psi}(\tau) = [\mathcal{H} + \mathcal{H}_{\rm int}]\boldsymbol{\Psi}(\tau),
\end{equation}
where $\bm{\Psi}(\tau)$ is a vector in the extended (Fourier coefficient) space, $\mathcal{H}$ is the extended space Hamiltonian (of the unperturbed system), defined as in Eq.~(\ref{eq:EZFloquet}), and $\mathcal{H}_{\rm int}$ is the extended space representation of the perturbation $H_{\rm int}$. 
Importantly, as noted above, $\mathcal{H}_{\rm int}$ is diagonal in Fourier harmonics, and takes the form $\mathcal{H}_{\rm int}^{(m,m')} = H_{\rm int} \delta_{mm'}$.
 Here the superscript $(m, m')$ indicates the block coupling Fourier component sectors $m$ and $m'$.

To see the physical significance of the auxiliary time evolution in extended space, consider the system in the absence of the perturbation, $H_{\rm int} = 0$, and let $\boldsymbol{\Psi}_n(\tau = 0) = \boldsymbol{\varphi}_n$, corresponding to a Floquet state with quasienergy $\epsilon_n$ as in Eq.~(\ref{eq:EZFloquet}): $\mathcal{H}\bm{\varphi}_n = \epsilon_n \bm{\varphi}_n$. 
Under the auxiliary time evolution in Eq.~(\ref{eq:EZEvolution}), the vector $\boldsymbol{\Psi}_n$ evolves to $\boldsymbol{\Psi}_n(\tau) = e^{-i\epsilon_n \tau/\hbar}\boldsymbol{\varphi}_n$.
Using the projector $\mathcal{P}(\omega\tau)$ defined above Eq.~(\ref{eq:EZ2Phys}) 
to carry out the mapping from extended space back to the physical Hilbert space as in Eq.~(\ref{eq:EZ2Phys}), we obtain
\begin{eqnarray}
\nonumber \mathcal{P}(\omega\tau) \bm{\Psi}_n(\tau) &=& e^{-i\epsilon_n\tau/\hbar} \sum_m e^{-im\omega\tau}\Ket{\phi_n^{(m)}}\\
 &=& \Ket{\psi_n(\tau)}.
\end{eqnarray}
Thus the physical time evolution of each Floquet state is generated by auxiliary time evolution in extended space (with constant extended space Hamiltonian $\mathcal{H}$) followed by projection to the physical Hilbert space via $\mathcal{P}(\omega\tau)$, and finally setting $\tau = t$ to obtain the state at time $t$.
 By linearity of Eq.~(\ref{eq:EZEvolution}), the same arguments can be applied to the evolution of any superposition of Floquet states.

Noting that Eq.~(\ref{eq:EZEvolution}) defines a Schr\"{o}dinger-like problem with a constant
perturbation, a textbook derivation of Fermi's golden rule applied directly to the extended space
evolution of $\bm{\Psi}(\tau)$ would directly yield Eq.~(\ref{eq:W_fi}). This association is
nontrivial, as the overlap $\bm{\Psi}^\dagger_1(t)\bm{\Psi}_2(t)$ between two Fourier coefficient vectors
$\bm{\Psi}_1(\tau = t)$ and $\bm{\Psi}_2(\tau = t)$ is generally {\it not} equal to the overlap between the
corresponding physical states $\Ket{\psi_1(t)}$ and $\Ket{\psi_2(t)}$. Rather, the correct relation
between the overlaps is
$\Amp{\psi_1(t)}{\psi_2(t)}~=~\bm{\Psi}_1^\dagger(t)\mathcal{P}^\dagger(\omega t)
\mathcal{P}(\omega t)\bm{\Psi}_2(t)$. 

To characterize the action $\mathcal{P}(\omega \tau)$, it is helpful to expand the auxiliary-time-evolved Fourier component vector $\bm{\Psi}(\tau)$ in the basis of eigenvectors of $\mathcal{H}$, $\{\bm{\varphi}_n\}$:
\be
\label{eq:EZ_expansion}  \bm{\Psi}(\tau) = \sum_n c_n(\tau) e^{-i\epsilon_n \tau}\bm{\varphi}_n.
\ee
Generically, under the projection $\mathcal{P}(\omega\tau)$ from extended space to the physical Hilbert space, multiple components of the expansion of $\bm{\Psi}(\tau)$ in Eq.~(\ref{eq:EZ_expansion}) may map to the same Floquet state due to the redundancy described in Eq.~(\ref{eq:FloquetRedundancy}) and the surrounding text.
Due to this possibility, the coefficients $\{c_n(\tau)\}$ may not be simply related to the amplitudes of the physical state $\Ket{\psi(t)}$ expanded in the basis of unperturbed Floquet eigenstates.
Crucially, however, the conservation of (quasi)energy that emerges at long times ensures that $\bm{\Psi}(\tau)$ can be decomposed as a superposition of Fourier coefficient vectors with nearly equal quasienergies: $\lim_{\tau\to \infty} c_n(\tau) = 0$ for $\epsilon_n \neq \epsilon_i$, where we recall that $\epsilon_i$ is the quasienergy of the initial state, evaluated for the unperturbed (extended space) Hamiltonian $\mathcal{H}$.
This in particular implies that all components of $\lim_{\tau\to\infty} \bm{\Psi}(\tau)$ map to {\it physically-distinct} (unperturbed) Floquet eigenstates $\mathcal{P}(\omega \tau)e^{-i\epsilon_n \tau}\bm{\varphi}_n = \Ket{\psi_n(t = \tau)}$ under the action of $\mathcal{P}(\omega\tau)$.
Therefore, in the long-time limit where the transition rate is evaluated, the squares of the coefficients in Eq.~(\ref{eq:EZ_expansion}), $|c_n(\tau)|^2$, directly give the probabilities of finding the physical state of the system in Floquet state $\Ket{\psi_n(t)}$.
Similar considerations apply for $\bm{\Psi}(t)$ that would be obtained using perturbation theory in $\mathcal{H}_{\rm int}$, as used in the derivation of Fermi's Golden Rule in extended space, Eq.~(\ref{eq:W_fi}). 

Note that we derived the transition rate $W_{fi}$ by considering the transition probability $P^{(1)}_{fi}(NT)$ at integer multiples of the driving period [see Eq.~(\ref{eq:P_def})].
However, the argument above shows that, at late times, the transition probability can be computed from the extended space evolution, even for non-integer multiples of the driving period [provided that $H_{\rm int}$ is weak enough that $W_{fi} \ll \omega$, see comment below Eq.~(\ref{eq:W_fi_final})].
Thus the specialization to stroboscopic times was performed without loss of generality.\\

\noindent {\it Example: Scattering rate between Floquet-Bloch states due to spontaneous phonon
emission}.--- To illustrate the utility of Eq.~(\ref{eq:W_fi}), we now discuss how to use the extended space
formalism to analyze the scattering rate for an electron in a Floquet-Bloch band due to its
coupling with a bosonic bath of phonons.
The noninteracting part of the Hamiltonian is quadratic in
both the electron and phonon creation and annihilation operators,
\begin{equation}
H(t)=H_{\rm el}(t)+\sum_\mathbf{q} \hbar \omega_\mathbf{q} \hat{b}^\dagger_{\mathbf{q}}\hat{b}_{\mathbf{q}}.
\label{eq: e-ph nonint}
\end{equation}
For simplicity, we consider a quadratic Hamiltonian for spinless electrons in a single band,
\begin{equation}
H_{\rm el}(t)=\sum_{\mathbf{k}} \hat{ c}^\dagger_{\mathbf{k}} H_\mathbf{k}(t) \hat{c}_\mathbf{k}.
\label{eq: el}
\end{equation}
The discussion below can be easily generalized to electronic dispersions containing multiple bands.
We consider an electron-phonon interaction of the typical form
 \begin{equation}
H_{\rm int}=\sum_{\mathbf{k},\mathbf{q}} U(\vec{q}) \hat{c}^\dagger_{\vec{k}} \hat{c}_{\vec{k}+\vec{q}}(\hat{b}^\dagger_{\mathbf{q}} + \hat{b}_{-\vec{q}})\ +\ \rm {h.c.}, \label{eq:HamElBos} \end{equation}
where the matrix element $U(\vec{q})$ characterizes the strength and form of the interaction.

For illustration, we consider a single electron in an otherwise empty Floquet band, and a zero-temperature phonon bath.
This situation will allow us to illustrate the intrinsic rates for spontaneous transitions between Floquet-Bloch states.
Specifically, we use Eq.~(\ref{eq:W_fi}) to calculate the rate for an electron initially in a Floquet state $\Ket{\psi_i(t)}$ with momentum $\mathbf{k}$ to scatter to a Floquet state $\Ket{\psi_f(t)}$ with final momentum $\mathbf{k}'$, while emitting a phonon with momentum $\vec{q} = \vec{k}-\vec{k}'$.

As described above, we translate the problem to extended space where the Hamiltonian $H_{\rm tot}(t) = H(t) + H_{\rm el-ph}$ [see Eqs.~(\ref{eq: e-ph nonint})-(\ref{eq:HamElBos})] is represented by the Fourier-space matrix $\mathcal{H}_{\rm tot} = \mathcal{H} + \mathcal{H}_{\rm int}$.
We associate a complete basis of Floquet states of the noninteracting Hamiltonian, Eq.~(\ref{eq: e-ph nonint}), with eigenvectors of $\mathcal{H}$ in the extended space, with quasienergies within a single (the ``first'') Floquet-Brillouin zone, $\varepsilon_{\rm min} <\varepsilon<\varepsilon_{\rm min}+\hbar\omega$.
Importantly, the unperturbed Hamiltonian $H(t)$ (and hence $\mathcal{H}$ in extended space) acts on states in the combined Hilbert space of electrons and phonons.
Therefore each unperturbed quasienergy $\epsilon_{\vec{k}, \{n_\vec{q}\}} = \mathcal{E}^{({\rm el})}_{\vec{k}} + \mathcal{E}^{({\rm ph})}_{\{n_\vec{q}\}}$
consists of a sum of the electronic quasienergy $\mathcal{E}^{({\rm el})}_{\vec{k}}$, which is an eigenvalue of $\mathcal{H}_{\rm el}$ [the extended space representation of $H_{\rm el}(t)$], and the phonon (quasi)energy $\mathcal{E}^{({\rm ph})}_{\{n_\vec{q}\}} = \sum_{\vec{q}} \hbar \omega_{\vec{q}} n_{\vec{q}}$, where $n_{\vec{q}}$ is the occupation number of mode $\vec{q}$.
Crucially, under this convention, in order for all {\it total} quasienergies to fall within the first Floquet-Brillouin zone, the separate electron and phonon contributions $\mathcal{E}^{({\rm el})}_{\vec{k}}$ and $\mathcal{E}^{({\rm ph})}_{\{n_\vec{q}\}}$ generically {\it do not} fall within the first Floquet-Brillouin zone (it is only their sum that does).
The value of the total phonon (quasi)energy $\mathcal{E}^{({\rm ph})}_{\{n_\vec{q}\}}$ thereby determines which extended space representation of the electronic state $\Ket{\psi_i(t)}$ is needed, using the freedom expressed in Eq.~(\ref{eq:FloquetRedundancy}), in order to yield a total quasienergy within the selected zone.

For the case of a zero temperature phonon bath that we consider, the quasienergy $\varepsilon_i$ of the initial state in Eq.~(\ref{eq:W_fi}) consists solely of that of the electron: $\varepsilon_i=\mathcal{E}^{({\rm el})}_i$
(recall that $\mathcal{E}^{({\rm el})}_i$ is an eigenvalue of $\mathcal{H}_{\rm el}$), with $\varepsilon_{\rm min} <\mathcal{E}^{({\rm el})}_i<\varepsilon_{\rm min}+\hbar\omega$.
The quasienergy of the final state is a sum of the electron and (emitted) phonon contributions, $\varepsilon_f=\mathcal{E}^{({\rm el})}_f + \hbar\omega_{\vec{q}}$.
Importantly, the requirement that the quasienergy is conserved in Eq.~(\ref{eq:W_fi}), i.e., $\varepsilon_f=\mathcal{E}^{({\rm el})}_f+\hbar\omega_\mathbf{q}$, implies that $\mathcal{E}^{({\rm el})}_f$ may not necessarily remain within the first Floquet-Brillouin zone.
Specifically, we may find that $\mathcal{E}^{({\rm el})}_f$ lies in the interval
$\varepsilon_{\rm min} + m\hbar\omega \le \mathcal{E}^{({\rm el})}_f < \varepsilon_{\rm min}+(m+1)\hbar\omega$ for nonzero integer $m$.
Such transitions, in which an electron is scattered between different Floquet-Brillouin zones, are referred to as ``Floquet-Umklapp'' processes.

To further illustrate the nature of Floquet-Umklapp processes, consider the case $m=-1$.
Here the electron spontaneously ``falls'' out the bottom of the Floquet-Brillouin zone from which it started, landing in a final state in the zone below.
In this sense, the many redundant copies of the Floquet spectrum in extended space (e.g., as depicted in Fig.~\ref{fig:FloquetSpectrum}a) may act like additional bands in the spectrum.
Importantly, if the quasienergy of the final electronic state $\Ket{\psi_f(t)}$ is ``folded back'' into the first Floquet-Brillouin zone [via Eq.~(\ref{eq:FloquetRedundancy})], it may appear as if the electronic quasienergy has spontaneously {\it increased} within the zone.
For this reason, Floquet-Umklapp processes are an important source of ``quantum heating'' in Floquet systems~\cite{Dykman2011}.

To assess the rates of Floquet-Umklapp processes and the significance of the heating they may produce, it is necessary to evaluate the transition matrix element $|\bm{\varphi}_f^\dagger\,\mathcal{H}_{\rm int}\,\bm{\varphi}_i|^2$, where $\bm{\varphi}_i$ and $\bm{\varphi}_f$ are the extended space Fourier component vectors describing the (tensor product) initial and final states of the electrons and phonons, together.
When an electron changes Floquet-Brillouin zones, this corresponds to the transformation in Eq.~(\ref{eq:FloquetRedundancy}): a shift of the electronic quasienergy by $m$ zones shifts all Fourier harmonics in the corresponding vector $\bm{\varphi}$ by $m$.
The matrix element $|\bm{\varphi}_f^\dagger\,\mathcal{H}_{\rm int}\,\bm{\varphi}_i|^2$ is only significant if both $\bm{\varphi}_i$ and $\bm{\varphi}_f$ have their support in the same region of Fourier harmonic space (see Sec.~\ref{sec:Trunc}).
This leads to a suppression of the rates in some (but not all) Floquet-Umklapp scattering processes. Similar considerations apply to scattering processes arising from interactions of electrons with the electromagnetic environment,  electron-electron interactions, and coupling to external leads.
For a detailed discussion and calculation of scattering rates for Floquet-Bloch states, see, e.g., Refs.~\onlinecite{Dehghani2014, Seetharam2015, Iadecola2015, Bilitewski2015a, Genske2015, Esin2018,Seetharam2019}.


\section{Gap opening due to circularly polarized driving fields}\label{sec:GapOpening}
In this section we provide an explicit calculation to demonstrate how a circularly polarized driving
 field can open a gap at the Dirac point in the massless Dirac model of Eq.~(1) in Ref.~\cite{RudnerLindnerReview}; see also Refs.~\cite{Oka2009,Kitagawa2010}.
\begin{figure}[b]
\includegraphics[width=\columnwidth]{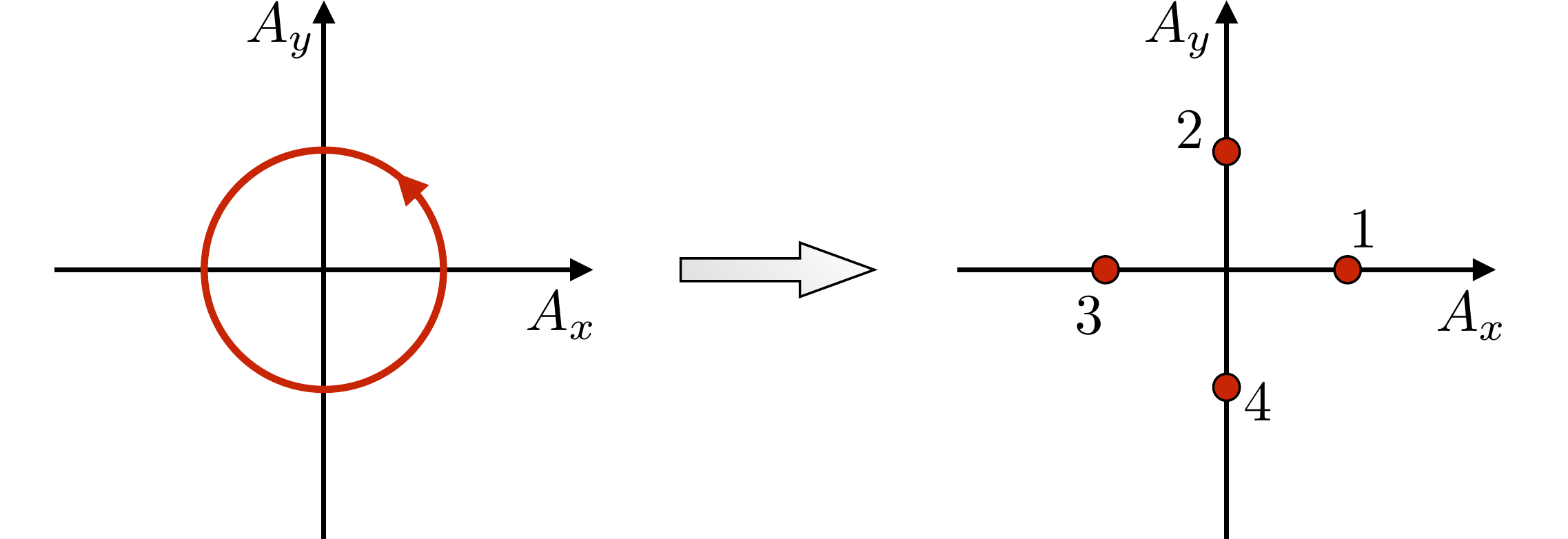}
\caption{To demonstrate Floquet gap opening at a Dirac point [Eq.~(1) of Ref.~\cite{RudnerLindnerReview}t], we consider a piecewise-constant driving protocol that mimics the effect of a circularly polarized field.
Rather than continuously rotating in the $xy$-plane, the vector potential with strength $A_0$ is taken to point in sequence along $x$, $y$, $-x$, and $-y$.
In each step, the field is applied for a duration $T/4$, such that the total driving period is $T \equiv 2\pi/\omega$.
The Floquet evolution operator $U(T)$ can be found exactly within this model.}
\label{fig:FourStepDrive}
\end{figure}
Rather than considering a continuously rotating circularly polarized harmonic driving field, we consider a four-step, piecewise-constant approximation to the circularly polarized drive as depicted in Fig.~\ref{fig:FourStepDrive}.
In each step $n$, the vector potential is held constant for a time $T/4$ with magnitude $A_0$ along the direction $\hat{\vec{e}}_n$ corresponding to point $n$ in the figure.
For each $\vec{k}$, the corresponding Bloch Hamiltonian for step $n$ is given by $H_n(\vec{k}) = v[\hbar\vec{k} - eA_0 \hat{\vec{e}}_n]\cdot\boldsymbol{\sigma}$.

The four-step drive maintains the chirality of the continuously rotating field, while allowing for a simple exact solution in the two-band model of Eq.~(1) of Ref.~\cite{RudnerLindnerReview}.
To obtain the effective Hamiltonian, we calculate the Floquet operator $U(\vec{k}, T) = U_4(\vec{k})U_3(\vec{k})U_2(\vec{k})U_1(\vec{k})$, where $U_n(\vec{k}) = e^{-i H_n(\vec{k})T/(4\hbar)}$.
For each $n$ and $\vec{k}$, $U_n(\vec{k})$ can be obtained simply by exponentiating Pauli matrices.

To illustrate gap opening we focus on the Dirac point and evaluate $U(\vec{k}~=~0, T)~=~e^{-i \phi \sigma_y} e^{-i \phi \sigma_x} e^{i \phi \sigma_y} e^{i \phi \sigma_x}$, with $\phi = evA_0 T/(4\hbar)$.
In the high frequency (short period, $T$) limit $\phi \ll 1$, the exponentials can be expanded to obtain
\begin{eqnarray}
 U(\vec{k} = 0, T) 
\label{eq:commutator} &=& \bm{1} + \phi^2 [\sigma_x, \sigma_y] + \mathcal{O}(\phi^3)\\
&=& \bm{1}+ 2i\phi^2 \sigma_z \approx e^{2i\phi^2\sigma_z}.
\end{eqnarray}
Using $U(\vec{k}=0, T) = e^{-i H_{\rm eff}(\vec{k}=0) T/\hbar}$, and $T = 2\pi/\omega$, we find $H_{\rm eff}(\vec{k} = 0) = \tilde{\Delta} \sigma_z$ with $\tilde{\Delta} = -\pi (evA_0)^2/(4\hbar\omega)$.

As this explicit calculation shows, the chiral driving field induces a gap at the Dirac point [cf.~Eq.~(1) of Ref.~\cite{RudnerLindnerReview}], with a magnitude that grows quadratically with the amplitude (linearly with the intensity) of the drive.
A very similar result is obtained for the case of the continuously rotating field, with a different prefactor.
The $\sigma_z$ term in the effective Hamiltonian arises due to the fact that $H(\vec{k}, t)$ does not commute with itself at different times.
Reversing the handedness of the drive would reverse the sign of the commutator in Eq.~(\ref{eq:commutator}), and hence also reverse the sign of the induced gap.

\section{Vanishing of the winding number $\nu_1$ for continuous time evolution}\label{sec:WindingVanish}
As stated in Sec.~II of Ref.~\cite{RudnerLindnerReview}, the winding number $\nu_1$ that counts the net winding of all Floquet bands around the quasienergy zone must vanish for any continuous evolution generated by a local, bounded Hamiltonian.
To see this, it is helpful to consider how the spectrum (band structure) of the evolution operator $U_{\rm 1D}(k, t) = \mathcal{T}e^{-(i/\hbar)\int_0^t dt' H_{\rm 1D}(k, t')}$ builds up as a function of time, $t$.
At $t = 0$, the evolution operator is the identity, $U_{\rm 1D}(k, t = 0) = {\bf 1}$.
At a small time $\delta t > 0$, the spectrum of $U_{\rm 1D}(k, \delta t)$ reflects the band structure of the instantaneous Hamiltonian, $H_{\rm 1D}(k, t = 0)$; crucially, the eigenvalues of $U_{\rm 1D}(k, \delta t)$ are periodic in $k$ and do not exhibit any nontrivial winding when $k$ traverses the Brillouin zone.
As time advances, the spectrum remains periodic in $k$ for all times, and by continuity cannot suddenly develop a net winding.
Therefore the spectrum of the Floquet operator $U_{\rm 1D}(k, T)$ cannot host a net winding of all of its bands; thus $\nu_1 = 0$.

\section{Definition of the time-averaged spectral function}\label{sec:TimeAvgSpec}
In Sec.~IV.A of Ref.~\cite{RudnerLindnerReview} (see in particular Fig.~3 and Fig.~\ref{fig:FloquetSpectrum}b herein) we introduced the ``time-averaged spectral function'' as a helpful tool for visualizing how mesoscopic transport occurs in Floquet systems.
In this section we give a formal definition of this quantity in terms of the system's retarded single-particle Floquet Green's function.
For completeness we review some basic features of Green's functions in Floquet systems, and discuss key similarities and differences between driven and non-driven systems (focusing on the non-interacting case).
Throughout this section we set $\hbar = 1$.

\subsection{Formal definitions}
We consider a Floquet system governed by a time-periodic Hamiltonian $H(t)$ satisfying $H(t + T) = H(t)$.
For a time-dependent Hamiltonian, we recall that the transformation to the Heisenberg picture is defined by the time evolution operator $U(t, t_0) = \mathcal{T}e^{-i\int_{t_0}^t dt' H(t')}$, which propagates the system forward in time from time $t_0$ to time $t$.
(Here $t_0$ is an arbitrary reference time, with respect to which the transformation is defined.)
The (time-independent) state of the system in the Heisenberg picture, $\ket{\psi}_H$, is obtained from the Schr\"{o}dinger picture state $\ket{\psi(t)}$ via $\ket{\psi}_H = U^\dagger(t, t_0)\ket{\psi(t)}$.
Importantly, a generic {\it time-dependent} operator $A(t)$ in the Schr\"{o}dinger picture transforms as
\begin{equation}
\label{eq:HeisenbergOp}A_H(t) = U^\dagger(t, t_0) A(t) U(t, t_0).
\end{equation}

Below we will work within the formalism of second quantization.
For brevity we will neglect spin throughout; the extension to include spin is straightforward.
We define the fermionic (``field'') creation operator $\Psi^\dagger(\vec{r})$, that creates a particle at position $\vec{r}$, and the annihilation operator $\Psi(\vec{r})$, that removes a particle from position $\vec{r}$.
These operators obey usual fermionic anticommutation relations, $\{\Psi(\vec{r}), \Psi^\dagger(\vec{r}')\} = \delta(\vec{r}-\vec{r}')$.

In the position representation, the single-particle retarded Green's function~\cite{Bruus2004} is defined 
as
\be
\label{eq:GR}G^R(\vec{r}t; \vec{r}'t') = -i\theta(t - t')\Avg{\{\Psi_H(\vec{r},t), \Psi^\dagger_H(\vec{r}',t') \}},
\ee
where the average is taken with respect to the many-body state of the system.
Here, the Heisenberg picture creation and annihilation operators $\Psi_H(\vec{r}, t)$ and $\Psi^\dagger_H(\vec{r}', t')$ are obtained from $\Psi(\vec{r})$ and $\Psi^\dagger(\vec{r}')$, respectively, using Eq.~(\ref{eq:HeisenbergOp}).
The retarded Green's function contains important information about the modes in which particles may propagate in the system, and will be the central object in the analysis below.

Similar to the case in equilibrium, for non- or weakly-interacting Floquet systems it is convenient to evaluate the Green's function in Eq.~(\ref{eq:GR}) in the basis of the single-particle Floquet states (``modes'') of the system in the absence of interactions, $\{\Ket{\psi_\nu(t)}\}$.
The Floquet modes are defined in the Schr\"{o}dinger picture as $\Ket{\psi_\nu(t)} = e^{-i\epsilon_\nu(t - t_0)}\Ket{\phi_\nu(t)}$, with $\Ket{\phi_\nu(t+T)} = \Ket{\phi_\nu(t)}$.
Here $\epsilon_\nu$ is the single-particle quasienergy associated with mode $\nu$.
In the position representation, mode $\nu$ is described by the time-periodic wave function $\phi_\nu(\vec{r}, t) = \Amp{\vec{r}}{\phi_\nu(t)}$.

In the {\it Schr\"{o}dinger picture}, we define the Floquet state creation operator at time $t$ as
\begin{equation}
c^\dagger_\nu(t) = \int d\vec{r}\, \phi_\nu(\vec{r},t) \Psi^\dagger(\vec{r}).
\label{eq: floquet creation}
\end{equation}
Thus, $c^\dagger_\nu(t)$ creates a particle in Floquet state $\nu$ at time $t$. [Note that we omit
the quasienergy phase factor from the definition of $c^\dagger_\nu(t)$.] The annihilation operator
$c_\nu(t)$ is defined analogously through Hermitian conjugation.

Crucially, for a {\it non-interacting system}, the Heisenberg operators $c^\dagger_{\nu, H}(t)$ and $c_{\nu,H}(t)$, defined via Eq.~(\ref{eq:HeisenbergOp}), take on a simple form in terms of the corresponding operators at the reference time $t_0$:
\begin{eqnarray}
\nonumber c^\dagger_{\nu, H}(t) &=& e^{i\epsilon_\nu(t-t_0)}c^\dagger_\nu(t_0)\\
\label{eq:Heisenberg_c} c_{\nu, H}(t) &=& e^{-i\epsilon_\nu(t-t_0)}c_\nu(t_0).
\end{eqnarray}
To see why Eq.~(\ref{eq:Heisenberg_c}) is true, note that the transformation to the Heisenberg picture corresponds to evolving backwards in time from $t$ to $t_0$: if a particle is created in Floquet mode $\nu$ at time $t$, then by its definition as a Floquet state it will evolve back to the corresponding mode $\nu$ at time $t_0$ under this transformation.
More formally, this property is reflected in the fact that in first quantized representation the evolution operator $U(t, t_0)$ for a single particle is diagonal with respect to the Floquet states, and can be written as $U(t, t_0) = \sum_\nu e^{-i\epsilon_\nu(t-t_0)}\Ket{\phi_\nu(t)}\Bra{\phi_\nu(t_0)}$.

Using Eq.~(\ref{eq:Heisenberg_c}), we obtain a crucial relation that we will use below for the evaluation of the retarded Green's function for non-interacting Floquet systems:
\be
\label{eq:FloquetCommutation} \{c_{\nu,H}(t), c^\dagger_{\nu',H}(t')\} = e^{-i\epsilon_\nu (t-t')}\delta_{\nu\nu'}.
\ee
To obtain this result, we used the orthogonality of the single-particle Floquet states at equal times: $\Amp{\phi_\nu(t_0)}{\phi_{\nu'}(t_0)} = \delta_{\nu\nu'}$.
Note the similarities between the relations in Eqs.~(\ref{eq:Heisenberg_c}) and (\ref{eq:FloquetCommutation}) and the corresponding relations for the Heisenberg picture creation and annihilation operators in a non-driven, non-interacting system.

\subsection{Properties of the Floquet retarded\\ Green's function}
In equilibrium, with or without interactions, the single-particle retarded Green's function [Eq.~(\ref{eq:GR})] has many useful properties.
For example $G^R(t,t')$ (with position or orbital indices suppressed) depends on $t$ and $t'$ only through the time difference $t - t'$.
After transforming to frequency space, the spectral function $A(\omega) = -\frac{1}{\pi} {\rm Im} [ G^R(\omega)]$ carries information about the energies of the characteristic single-particle modes of the system. 
The trace (i.e., sum over all states) of the spectral function yields the density of states of the system.
Importantly, $A(\omega)$ is a positive semi-definite matrix (in the space of single particle orbitals), and for non-interacting systems does not depend on the state of the system.

In contrast, for a driven system, $G^R(t, t')$ is a function of both $t$ and $t'$, not only the
time difference. Nonetheless, analogous spectral information to that familiar from equilibrium
systems can be obtained for Floquet systems (see
Refs.~\cite{Platero2004, Kohler2005, Tsuji2008,FoaTorresMultiTerminal,Usaj2014,Qin2017, Kalthoff2018,Uhrig2019}), provided
that the system is non-interacting and/or that it is in a time-periodic steady state. Under these
conditions, $G^R(t, t')$ is periodic in the average (or ``center of mass'') time $\bar{t} =
\frac12(t + t')$ with period $T$: shifting both $t$ and $t'$ by $T$ leaves the state of the system
invariant. As shown rigorously in Ref.~\cite{Uhrig2019}, a positive spectral density analogous to
that in equilibrium is then obtained by averaging $G^R(\bar{t} + \frac12 \tau,
\bar{t}-\frac12\tau)$, with $\tau = t - t'$, over one full period in the average time $\bar{t}$,
and then Fourier transforming with respect to the time-difference, $\tau$.

\begin{widetext}
\subsection{Evaluation of the Floquet retarded Green's function for a non-interacting system}
In the remainder of this section we illustrate the properties above with an explicit calculation of the retarded Green's function for a {\it non-interacting} Floquet system.
Returning to Eq.~(\ref{eq:GR}), we express the field operators $\Psi^\dagger_H(\vec{r}',t')$ and $\Psi_H(\vec{r},t)$ in terms of Floquet state creation and annihilation operators.
There is some freedom in how to do this; the most fruitful way is to write $\Psi^\dagger_H(\vec{r}',t') = U^\dagger(t', t_0) \Psi^\dagger(\vec{r}') U(t', t_0)$ and $\Psi(\vec{r}, t) = U^\dagger(t, t_0) \Psi(\vec{r}) U(t, t_0)$, and then to express $\Psi(\vec{r})$ and $\Psi^\dagger(\vec{r}')$ in terms of the (Schr\"{o}dinger picture) operators $\{c_\nu(t)\}$ and $\{c^\dagger_{\nu'}(t')\}$.
{Using the inverse transformation of Eq.~(\ref{eq: floquet creation}), this yields $ \Psi_H(\vec{r},t) = \sum_\nu \phi_\nu(\vec{r}, t) c_{\nu, H}(t)$ and $\Psi^\dagger_H(\vec{r}',t') = \sum_\nu \phi^*_\nu(\vec{r}', t') c^\dagger_{\nu', H}(t)$}.
Substituting into Eq.~(\ref{eq:GR}) and using Eq.~(\ref{eq:FloquetCommutation}), we obtain
\be
\label{eq:GR_non}{G_0^R(\vec{r}t; \vec{r}'t') = -i\theta(t-t')\sum_\nu \phi_\nu(\vec{r},t)\phi^*_\nu(\vec{r}', t') e^{-i\epsilon_\nu (t-t')}.}
\ee
Here we use the subscript 0 to emphasize that the result holds only for non-interacting systems.

Due to the periodicity of the wave functions $\{\phi_\nu(\vec{r},t)\}$, we express {$\phi_\nu(\vec{r},t)\phi^*_\nu(\vec{r}', t')$} as a Fourier series in terms of harmonics $\{\phi_\nu^{(m)}(\vec{r})\}$ of the drive frequency, $\omega$: {$\phi_\nu(\vec{r},t)\phi^*_\nu(\vec{r}', t')~=~\sum_{mm'} e^{-im\omega (\bar{t} + \tau/2)}e^{im'\omega (\bar{t}-\tau/2)} \phi_\nu^{(m)}(\vec{r})\big[\phi_\nu^{(m')}(\vec{r}')\big]^*$}.
Inserting this expression into Eq.~(\ref{eq:GR_non}) and averaging with respect to $\bar{t}$ over one full period (for fixed $\tau$), we obtain the time-averaged single-particle retarded Green's function, $ \bar{G}_0^R(\vec{r}, \vec{r}'; \tau)$:
\bea
 \label{eq:GR_avg}\bar{G}_0^R(\vec{r}, \vec{r}'; \tau) &\equiv& \frac{1}{T}\int_{t_0}^{t_0 + T} d\bar{t}\,G_0^R(\vec{r, }\bar{t} + \tau/2; \vec{r}', \bar{t} - \tau/2)\\
\nonumber &=& {-i\theta(\tau)\sum_{\nu,m} e^{-i(\epsilon_\nu + m\omega)\tau}\phi_\nu^{(m)}(\vec{r})\big[\phi_\nu^{(m)}(\vec{r}')\big]^*.}
\eea

Taking the Fourier transform  $\tilde{G}_0^R(\vec{r}, \vec{r}'; \Omega) = \lim_{\eta \to 0^+}\int_{-\infty}^\infty d\tau\, e^{i(\Omega + i\eta)\tau}\bar{G}_0^R(\vec{r}, \vec{r}'; \tau)$,  we obtain
\be
\label{eq:GR_avg2}{\tilde{G}_0^R(\vec{r}, \vec{r}'; \Omega) = \lim_{\eta \to 0^+}\sum_{\nu,m} \frac{\phi_\nu^{(m)}(\vec{r})\big[\phi_\nu^{(m)}(\vec{r}')\big]^*}{\epsilon_\nu + m\omega - \Omega + i\eta}.}
\ee
Analogous to equilibrium, we define the ``time-averaged density of states'' via $\bar{\rho}_0(\Omega) = -\frac{1}{\pi} {\rm Tr}\, {\rm Im}[\tilde{G}_0^R(\Omega)]$:
\be
\label{eq:DOS} \bar{\rho}_0(\Omega) = {\sum_{\nu,m} A^{(m)}_\nu \delta(\epsilon_\nu + m\omega - \Omega)}, \quad A^{(m)}_\nu = \Amp{\phi_\nu^{(m)}}{\phi_\nu^{(m)}},
\ee
where $A^{(m)}_\nu$ 
captures the spectral weight of the $m$-th harmonic/sideband component $\ket{\phi_\nu^{(m)}}$ in the Floquet state $\ket{\psi_\nu(t)}$.
(Here $\ket{\phi_\nu^{(m)}}$ is the ket corresponding to the component $\phi^{(m)}_\nu(\vec{r})$ of the position-space Floquet state wave function defined above.)

As seen in Eq.~(\ref{eq:DOS}), the time-averaged density of states is comprised of delta-function peaks at frequencies corresponding to the quasienergies of the single-particle Floquet states of the system, plus or minus all integer multiples of the driving field photon energy, $\hbar \omega$.
The weight of each peak is governed by the spectral weight $A^{(m)}_\nu$ of the corresponding sideband component of the Floquet state.

To obtain the time-averaged spectral function, we single out the contribution from a given Floquet
eigenstate $\nu$: {$\mathcal{A}_\nu(\Omega) = \sum_m A^{(m)}_\nu \delta(\epsilon_\nu + m \omega -
\Omega)$}. The time-averaged spectral function captures the fact that the spectral weight of each
Floquet state is spread in frequency (energy) over several discrete harmonics. 
As displayed in Fig.~\ref{fig:FloquetSpectrum}b, the spectral function for graphene subjected to circularly polarized light shows, in highest weight, the original graphene (projected) band structure, interrupted by dynamically-induced band gaps at zero energy and at resonances where states in the original conduction and valence bands are separated by integer multiples of $\hbar\omega$.
Sidebands (of low spectral weight) appear as faint copies of these bands, shifted up and down by integer multiples of $\hbar\omega$.
As discussed in Ref.~\cite{RudnerLindnerReview}, the time-averaged spectral function thus provides a helpful way to resolve
how states at specific {\it energies} in the leads couple to Floquet states with given
quasienergies in the system \cite{FoaTorresMultiTerminal,Farrell2015,Kundu2013,Kundu2017}.

\section{Floquet-Kubo formula}\label{sec:FloquetKubo}
In this section we discuss the linear response of a Floquet system to the application of an
additional weak perturbing field, oscillating at a frequency $\Omega$ that we assume to be much
smaller than the driving frequency $\omega$. 
(For an extended discussion of linear response theory for Floquet systems, see Ref.~\onlinecite{Kumar2019}.)
The Hamiltonian, including the perturbation, is given
by $\hat H(t)=\hat H_0(t)+F(t)\hat B(t)$, where $F(t)$ is a general function of time, and $\hat
B(t)$ is an operator that may depend on time explicitly in the Schr\"{o}dinger picture, albeit in a
periodic manner with the same period as $\hat{H}_0$: $\hat B(t)=\hat B(t+T)$. The derivation given
in this section closely follows the derivation of the Kubo formula for equilibrium
systems~\cite{Bruus2004}. 
For a derivation of the Floquet-Kubo formula based on the $tt'$ formalism, see Ref.~\onlinecite{Wackerl2019}.
Throughout this section we will denote operators using hats to avoid
possible ambiguity about which objects are operators and which are simply real or complex numbers.

As a starting point, we assume that in the absence of the perturbing field the system has reached a time-periodic stationary state, described by a density matrix $\hat{\rho}_0(t)$, with $\hat{\rho}_0(t + T) = \hat{\rho}_0(t)$.
The change in the expectation value of a constant or time-periodic operator $\hat{A}(t)$ in response to the perturbing field $F(t)$ is given by
\begin{equation}
\delta \langle \hat A(t)\rangle = \int_{-\infty}^\infty dt'\, \tilde\chi(t,t') F(t'), \label{eq:linear response}
\end{equation}
where $\delta \langle \hat A(t)\rangle = \Tr \left[\hat\rho(t)\hat A(t)\right]-\Tr \left[\hat\rho_0(t)\hat A(t)\right]$.
Here,  $\hat\rho(t)$ is the state of the system in the presence of the perturbation. 
Analogous to the case in equilibrium, the retarded response function $\tilde\chi(t,t')$ is defined as
\begin{equation}
\tilde\chi(t,t')=-i\theta(t-t')\Tr\left\{\hat \rho_0(t_0) \left[\hat A_I(t),\hat B_I(t')\right]\right\},
\label{eq: response function}
\end{equation}
where the interaction picture operator $\hat{A}_I(t)$ is defined by the transformation
\begin{equation}
\hat A_I(t)=\hat{U}_0^\dagger(t,t_0)\hat A(t) \hat{U}_0(t,t_0),
\end{equation}
with  $\hat{U}_0(t,t_0)=\mathcal{T}e^{-i \int_{t_0}^t \hat H_0(t')dt'}$, where $\mathcal{T}$ denotes time
ordering. The operator $\hat B_I(t')$ is defined in a similar manner. Note that although the arbitrary reference time
$t_0$ appears explicitly in Eq.~(\ref{eq: response function}), the response function $\tilde\chi(t,t')$ is in fact independent of $t_0$.
This can be checked by using the definitions of $\hat\rho_0(t_0)$ and of the operators $\hat{A}_I(t)$ and $\hat{B}_I(t')$.

Since the system is assumed to be in a \textit{steady state}, the response function $\tilde{\chi}(t,t')$ is
invariant under shifting both $t$ and $t'$ by $T$, the driving period.  To see why, we note the
steady state condition implies that $\hat{\rho}_0(t_0)=\hat{\rho}_0(t_0-T)$. The time-periodicity of the
Hamiltonian implies $\hat{U}_0(t+T,t_0)=\hat{U}_0(t,t_0-T)$, and therefore
\begin{equation}
\begin{split}
\Tr\left\{\hat\rho_0(t_0) \left[\hat A_I(t+T),\hat B_I(t'+T)\right]\right\}
&= \Tr\left\{\hat\rho_0(t_0-T) U_0^\dagger(t_0,t_0-T)\left[\hat A_I(t),\hat B_I(t')\right]U_0(t_0,t_0-T)\right\}\\
&=\Tr\left\{\hat\rho_0(t_0) \left[\hat A_I(t),\hat B_I(t')\right]\right\}.
\end{split}
\end{equation}
Transforming variables using $\tau \equiv t - t'$ we see that the function $ \chi(\tau,t') \equiv \tilde{\chi}(t' + \tau,t')$, given by
\begin{equation}
\chi(\tau,t')=-i\theta(\tau)\Tr\left\{\rho_0(t_0) \left[\hat A_I(t'+\tau),\hat B_I(t')\right]\right\},
\end{equation}
is periodic in  $t'$ with period $T$ (for fixed $\tau$). 
Thus $\chi(\tau, t')$ has a mixed Fourier representation in terms of one continuous frequency
variable, $\Omega$, conjugate to $\tau$, and a discrete set of harmonics, $m$, which capture its
periodicity in $t'$:
\begin{equation}
\label{eq:chi_transf}\chi(t-t',t')=\sum_m \int_{-\infty}^\infty d\Omega\, e^{-i \Omega (t-t')}e^{-i m\omega t'} \chi^{(m)}(\Omega).
\end{equation}
Inserting Eq.~(\ref{eq:chi_transf}) into Eq.~(\ref{eq:linear response}) and taking the Fourier
transform of both sides with respect to $t$, we obtain
\begin{equation}
\delta \langle \hat A\rangle(\Omega)=\sum_m  \chi^{(m)}(\Omega) F(\Omega-m\omega).
\label{eq: freq response}
\end{equation}

Suppose that i) we probe the observable $\hat A(t)$ at frequencies $\Omega$ that are much smaller than the
drive frequency, $\omega$, and ii) the Fourier transform of the perturbing field, ${F}(\Omega')$, only has support at frequencies $\Omega'\ll\omega$.
(The latter captures our original assumption that the system is perturbed at a frequency much less than that of the drive.)
Under these conditions, the only significant term in  the sum in  Eq.~(\ref{eq: freq response}) corresponds to $m=0$, giving
\begin{equation}
\delta\langle  \hat{A}\rangle(\Omega) =\chi^{(0)}(\Omega)F(\Omega).
\label{eq: low freq response}
\end{equation}
Note that $\chi^{(0)}(\Omega)$ is just the Fourier transform of the ``time-averaged response function,''  $\bar{\chi}(\tau) =\frac{1}{T}\int_0^T dt'
\chi(\tau,t')$, with respect to the variable $\tau$.

\subsection{Floquet-Kubo Formula for the electrical conductivity}

We now focus on the response of an electromagnetically-driven electronic Floquet system to a
uniform probing AC electric field oscillating at a frequency
$\Omega$, where as before $\Omega\ll \omega$~\cite{TorresFloquetKubo2005,Oka2009}. The extension to finite wave vector probe fields can
be performed using a similar approach. The electromagnetic gauge potential is then a sum of two
terms, $\bA(t)=\bA_0(t)+\delta \bA(t)$, where $\bA_0(t)=\bA_0(t+T)$ describes the time-periodic
driving field (which we aim to include exactly), and $\delta \bA(t)$ describes the weak probe
field. To second order in $\delta \bA(t)$, the Hamiltonian is given by
\begin{equation}
\hat H(t)= \hat H(t)\Big|_{\delta  \bA(t)=0} +\ \frac{\partial \hat{H}(t)}{\partial \bA(t)}\Bigg|_{\delta  \bA(t)=0}\!\!\!\!\!\!\!\!\!\!\!\!\!\!\!\cdot\, \delta  \bA(t)
\ +\ \frac{\partial^2 \hat{H}(t)}{\partial A_\alpha(t) \partial A_\beta(t)}\Bigg|_{\delta  \bA(t)=0}\!\!\!\!\!\!\!\!\!\!\!\!\!\!\!\delta A_\alpha(t) \delta A_\beta(t),
\label{eq: ham 2nd order}
\end{equation}
where $\alpha, \beta = \{x, y, z\}$.
The current operator $\hat{\mathbf{J}}(t)=\left(\hat{J}_x(t),\hat{J}_y(t),\hat{J}_z(t)\right)$ is
given by
\begin{equation}
\hat{\mathbf{J}}(t)=-\frac{\partial \hat{H}(t)}{\partial \bA(t)}. 
\end{equation}

Importantly, due to the time-dependence of $\bA_0(t)$ in $\hat H(t)$, the current operator
$\hat{\vec{J}}(t)$ is time-dependent (in the Schr\"{o}dinger picture). In analogy to common
practice in equilibrium systems, we now define a ``paramagnetic current operator''
$\hat{\bJ}^{(p)}(t)$ and the ``kinetic'' operator $\hat{K}_{\alpha\beta}(t)$ as
\cite{MahanBook,Scalapino1993}:
\begin{equation}
\hat{\bJ}^{(p)}(t)=-\frac{\partial \hat{H}(t)}{\partial  \bA(t)}\Bigg|_{\delta  \bA(t)=0},\quad \hat{K}_{\alpha\beta}(t)=-\frac{\partial^2 \hat{H}(t)}{\partial A_\alpha(t) \partial A_\beta(t)}\Bigg|_{\delta  \bA(t)=0}.
\label{eq: def para dia}
\end{equation}

Using Eqs.~(\ref{eq: ham 2nd order})--(\ref{eq: def para dia}), the expectation value of the current, up to first order in $\delta \bA(t)$, can be written as a sum  of ``paramagnetic'' and ``diamagnetic'' contributions,
\begin{equation}
\langle\hat{J}_\alpha(t)\rangle=\langle \hat{J}_\alpha^{(p)}(t)\rangle + \sum_\beta\Tr \left[\hat{\rho}_0(t) \hat{K}_{\alpha\beta}(t)\right]\delta A_\beta(t).
\label{eq: para plus dia}
\end{equation}
For clarity we note that the expectation value $\langle\hat{J}_\alpha(t)\rangle \equiv
\Tr[\hat{\rho}(t)\hat{J}_\alpha(t)]$ of the time-dependent operator $\hat{J}_\alpha(t)$
is taken  with respect to the (full) state of the system $\hat\rho(t)$ at the same time, $t$. 
To obtain an equation valid within the
regime of linear response, in Eq.~(\ref{eq: para plus dia}) we expand the expectation value
$\langle\hat{J}^{(p)}_\alpha(t)\rangle$  up to linear order in the probe field $\delta\bA(t)$,
while we take the expectation value in the second term on the RHS 
with respect to the unperturbed state $\hat{\rho}_0(t)$.

The change in the paramagnetic current due to the probe field $\delta \bA(t)$ (relative to any time
periodic current that may flow in the steady state) is given by
\begin{equation}
\delta\langle \hat{J}_\alpha^{(p)}(t)\rangle =\int_{-\infty}^{\infty}dt'\chi_{\alpha\beta}(t-t',t') \delta A_\beta(t'),
\label{eq: para time}
\end{equation}
where
\begin{equation}
\chi_{\alpha\beta}(\tau,t')=-i\theta(\tau)\Tr\left\{\rho_0(t_0) \left[\hat{J}_{\alpha,I}^{(p)}(t'+\tau),\hat{J}_{\beta,I}^{(p)}(t')\right]\right\}.
\end{equation}
Taking the Fourier transform of Eq.~(\ref{eq: para time}) with respect to $t$, and using the
periodicity of $\chi_{\alpha\beta}(t-t',t')$ with respect to $t'$, we obtain
\begin{equation}
\delta \langle \hat{J}_\alpha^{(p)}\rangle(\Omega)=\int_{-\infty}^{\infty} dt' \sum_m e^{i(\Omega-m\omega)t'}\chi^{(m)}_{\alpha\beta}(\Omega)\delta A_\beta(t').
\label{eq: para omega_0}
\end{equation}
For the electromagnetic gauge potential, we choose a gauge for which $\bE(t)=-\frac{d}{dt}\bA(t)$.
Using integration by parts on Eq.~(\ref{eq: para omega_0}) to obtain a time derivative on $\delta
A_\beta(t)$ yields
\begin{equation}
\delta \langle \hat{J}_\alpha^{(p)}\rangle(\Omega)=\sum_m \frac{\chi^{(m)}_{\alpha\beta}(\Omega)}{i(\Omega-m\omega)}\delta E_\beta(\Omega-m\omega).
\label{eq: para omega}
\end{equation}
For electric fields that contain only low frequencies, $\Omega\ll\omega$, only the $m=0$
component contributes to Eq.~(\ref{eq: para omega}).

In the following we will assume that the gauge (probe) field $\delta\bA(t)$ appears only in the
quadratic part of the Hamiltonian, $\hat H^{(2)}(t)=\sum_{ij}  h_{ij}(t) \hat d^\dagger_i \hat
d_j$, where $\hat{d}^\dagger_i$ is an electronic creation operator with respect to a fixed,
time-independent basis. In terms of the basis of Floquet states, the quadratic part of the
Hamiltonian can be written as $\hat{H}^{(2)}(t)=\sum_{\bk} h_{\nu_1\nu_2}(\bk,t)
\hat{c}^\dagger_{\bk\nu_1}(t)\hat c_{\bk\nu_2}(t)$, where $\{\hat c^\dagger_{\bk\nu}(t)\}$ are
creation operators of electrons in single particle Floquet states, see Eq.~(\ref{eq: floquet
creation}), and $\nu$ is a Floquet band index. We then express the paramagnetic part of the current
operator as
\begin{equation}
\hat{J}^{(p)}_\alpha(t)=\sum_{\bk} j_{\alpha,\nu_1\nu_2}(\bk,t) \hat c^\dagger_{\bk\nu_1}(t)\hat c_{\bk\nu_2}(t),
\label{eq: para quad}
\end{equation}
where the matrix elements $j_{\alpha, \nu_1\nu_2}(\bk,t)$ are given by
\begin{equation}
\label{eq:j_def}j_{\alpha,\nu_1\nu_2}(\bk,t)=-\frac{\partial h_{\nu_1\nu_2}(\bk,t)}{\partial A_\alpha(t)}\Bigg|_{\delta  \bA(t)=0}=-\big\langle \phi_{\bk,\nu_1}(t)\big|\frac{\partial \check{h}(t)}{\partial  A_\alpha(t)}\big|\phi_{\bk,\nu_2}(t)\big\rangle\Big|_{\delta  \bA(t)=0},
\end{equation}
where $\check{h}(t)$ is the single-particle (first quantized) operator corresponding to the matrix
$h_{ij}(t)$.

We focus on the case of a {\it diagonal} steady state of the form
\begin{equation}
 \hat \rho_0(t)=\prod_{\bk\nu}[f_{\bk\nu}\hat c^\dagger_{\bk\nu}(t) \hat
c_{\bk \nu}(t)+(1-f_{\bk\nu})\hat c_{\bk\nu}(t) \hat c^\dagger_{\bk\nu}(t)],
\label{eq: steady rho}
\end{equation}
where  $f_{\bk\nu}$ is the population of the Floquet state created by $\hat c^\dagger_{\bk\nu}(t)$.
As a useful preliminary for the evaluation of the response function, we note that for a steady
state of the form in Eq.~(\ref{eq: steady rho}), the analogue of Eq.~(\ref{eq:Heisenberg_c}) for
the interaction picture yields
\begin{equation}
\Tr\left[\hat\rho_0(t_0)\hat{c}^\dagger_{I, \bk\nu}(t_1) \hat{c}_{I, \bk'\nu'}(t_2)\right]=f_{\bk\nu}e^{i\varepsilon_{\bk\nu}(t_1-t_2)}\delta_{\bk\bk'}\delta_{\nu\nu'}.
\label{eq: chi prelim}
\end{equation}

We now use the results and definitions above to evaluate the linear response conductivity of the
Floquet system.
Using Eq.~(\ref{eq: para quad})--Eq.~(\ref{eq: chi prelim}), the response
function $\chi_{\alpha\beta}(\tau,t')$ is evaluated to be
\begin{equation}
\chi_{\alpha\beta}(\tau,t')=  -i\theta(\tau)\sum_{\bk}\sum_{\nu_1\nu_2}e^{i(\varepsilon_{\bk \nu_1}-\varepsilon_{\bk \nu_2})\tau}  \left(f_{\bk\nu_1}-f_{\bk\nu_2}\right)  j_{\alpha, \nu_1\nu_2}(\bk,t' + \tau)j_{\beta, \nu_2\nu_1}(\bk,t').
\end{equation}\\

 Using the time-periodicity of $j_{\alpha,
\nu_1\nu_2}(\bk,t)$ [see definition in Eq.~(\ref{eq:j_def})], we expand $j_{\alpha,
\nu_1\nu_2}(\bk,t)=\sum_m e^{-i m \omega t}j^{(m)}_{\alpha, \nu_1\nu_2}(\bk)$. Averaging the time
variable $t'$ over one period and taking the Fourier transform with respect to $\tau$, we obtain
the time-averaged response function
\begin{equation}
\label{eq:chi0final}  \chi^{(0)}_{\alpha\beta}(\Omega)= \lim_{\eta \to 0^+}\sum_{\bk}\sum_{\nu_1\nu_2}\sum_m\frac{\left(f_{\bk\nu_1}-f_{\bk\nu_2}\right)j^{(m)}_{\alpha, \nu_1\nu_2}(\bk)j^{(-m)}_{\beta,\nu_2\nu_1}(\bk)}{{\Omega}-m\omega+(\varepsilon_{\bk \nu_1}-\varepsilon_{\bk \nu_2}) + i \eta}.
\end{equation}
Using the gauge $\bE(t)=-\frac{d}{dt}\bA(t)$ as before, we obtain a contribution from the
diamagnetic (second) term in Eq.~(\ref{eq: para plus dia}) to the total current $\langle
\hat{J_\alpha}\rangle (\Omega)\equiv\int_{-\infty}^{\infty} dt\, e^{i\Omega t}\langle
\hat{J_\alpha}(t)\rangle$ that is given by:
\begin{equation}
\int_{-\infty}^\infty dt\, e^{i\Omega t} \Tr\left[\hat\rho_0(t)\hat{K}_{\alpha\beta}(t)\right]\delta A_\alpha(t)=\sum_m\frac{ \mathcal{K}_{\alpha\beta}^{(m)}}{i(\Omega-m\omega)}\delta E_\alpha(\Omega+m\omega).
\label{eq: dia}
\end{equation}
In writing Eq.~(\ref{eq: dia}) we have used the time-periodicity of
$\mathcal{K}_{\alpha\beta}(t)=\Tr\left[\hat\rho_0(t)\hat{K}_{\alpha\beta}(t)\right]$ and expanded
$\mathcal{K}_{\alpha\beta}(t)=\sum_m \mathcal{K}^{(m)}_{\alpha\beta} e^{-im\omega t}$. Using this
expansion, Eq.~(\ref{eq: dia}) is obtained by integration by parts. Note that for electric fields
that contain only low frequency components $\Omega\ll\omega$, only the term $m=0$ in Eq.~(\ref{eq:
dia}) gives a contribution to the current. In this case, this second contribution to the current is
given by the time-average of $\mathcal{K}_{\alpha\beta}(t)$,
\begin{equation}
\mathcal{K}^{(0)}_{\alpha\beta}=\frac{1}{T}\int_0^T dt\, \mathcal {K}_{\alpha\beta}(t).
\end{equation}

Finally we obtain Ohm's law for the response to the probe field, $\delta \Avg{\hat{J}_\alpha}(\Omega) =
\sigma_{\alpha\beta}(\Omega) \delta E_\beta(\Omega)$: 
\begin{equation}
\sigma_{\alpha\beta}(\Omega)=\frac{\chi^{(0)}_{\alpha\beta}(\Omega)+\mathcal{K}^{(0)}_{\alpha\beta}}{i\Omega},\quad \textrm{(for $\Omega\ll\omega$)}.
\end{equation}
Note the similarity between this result and the analogous expression for the conductivity in equilibrium systems, as well the form of the response function $\chi^{(0)}_{\alpha\beta}(\Omega)$ in Eq.~(\ref{eq:chi0final}).

\end{widetext}

\section{Acknowledgements}
We thank all of our collaborators on FTI-related work, with whom we have had many stimulating
interactions. In particular, we acknowledge Erez Berg, Eugene Demler, Victor Galitski, Takuya
Kitagawa, Michael Levin and Gil Refael, with whom we began our journey in this field. N. L. acknowledges support from the
European Research Council (ERC) under the European Union Horizon 2020 Research and Innovation
Programme (Grant Agreement No. 639172), and from the Israeli Center of Research Excellence (I-CORE)
``Circle of Light''. M. R. gratefully acknowledges the support of the European Research Council
(ERC) under the European Union Horizon 2020 Research and
Innovation Programme (Grant Agreement No.678862), the Villum Foundation, and CRC 183 of the Deutsche Forschungsgemeinschaft. 

\bibliography{FloquetTopologicalInsulator_references}

\end{document}